\PassOptionsToPackage{table}{xcolor} % must appear before \documentclass if using acmart

\documentclass[manuscript]{acmart}
\usepackage[most]{tcolorbox}
\usepackage{colortbl}

\usepackage{booktabs}
\usepackage{multirow}
\usepackage{xcolor}
\usepackage{tabularx}

\usepackage{subcaption}
\usepackage{wrapfig}
\usepackage{caption}

\usepackage{xspace}

\usepackage{tikz}
\usepackage{soul}

\definecolor{lightgray}{gray}{0.7}
\definecolor{headergray}{gray}{0.9}
\newcommand{\Dim}[2]{{\sffamily\slshape\setul{1pt}{1pt}\setulcolor{#1}\ul{#2}}}

\DeclareRobustCommand{\Code}[2]{%
  \tikz[baseline=(X.base)]{
    \node[
      fill=#1,
      inner sep=1pt,
      rounded corners=1mm,
      anchor=base
    ] (X) {\color{black}\textit{#2}};
  }%
}

\definecolor{GrayFill}{RGB}{222, 222, 222}
\definecolor{GrayUl}{RGB}{197, 197, 197}
\newcommand{\gray}[1]{\Code{GrayFill}{#1}}
\newcommand{\Gray}[1]{\Dim{GrayUl}{#1}}

\definecolor{PeachFill}{RGB}{255, 207, 207}
\definecolor{PeachUl}{RGB}{249, 136, 116}

\definecolor{BlueFill}{RGB}{175, 216, 255}
\definecolor{BlueUl}{RGB}{143, 193, 255}
\newcommand{\blue}[1]{\Code{BlueFill}{#1}}
\newcommand{\Blue}[1]{\Dim{BlueUl}{#1}}

\definecolor{PinkFill}{RGB}{249, 207, 255}
\definecolor{PinkUl}{RGB}{247, 157, 205}

\definecolor{PurpleFill}{RGB}{234, 207, 255}
\definecolor{PurpleUl}{RGB}{210, 167, 255}
\newcommand{\purple}[1]{\Code{PurpleFill}{#1}}
\newcommand{\Purple}[1]{\Dim{PurpleUl}{#1}}

\definecolor{GreenFill}{RGB}{207, 255, 209}
\definecolor{GreenUl}{RGB}{158, 238, 162}

\definecolor{YellowFill}{RGB}{255, 221, 131}
\definecolor{YellowUl}{RGB}{243, 207, 128}
\newcommand{\yellow}[1]{\Code{YellowFill}{#1}}
\newcommand{\Yellow}[1]{\Dim{YellowUl}{#1}}

\newcommand{\purposecode}[1]{\gray{#1}}
\newcommand{\Purpose}[1]{\Gray{#1}}

\newcommand{\nodecode}[1]{\blue{#1}}
\newcommand{\Node}[1]{\Blue{#1}}

\newcommand{\relationcode}[1]{\yellow{#1}}
\newcommand{\Relation}[1]{\Yellow{#1}}

\newcommand{\errorcode}[1]{\purple{#1}}
\newcommand{\Error}[1]{\Purple{#1}}

\AtBeginDocument{%
  }

\setcopyright{acmlicensed}
\copyrightyear{2026}
\acmYear{2026}
\acmDOI{XXXXXXX.XXXXXXX}
\acmConference[Conference acronym 'XX]{Make sure to enter the correct
  conference title from your rights confirmation email}{MMMM DD--DD,
  2026}{Woodstock, NY}
\acmISBN{978-1-4503-XXXX-X/2026/06}

\begin{document}

\title{Planning on Paper: Problem Decomposition with Diagrams in Introductory Computing}

%[Exploring Problem Decomposition and Planning in CS1]
%Patterns and Errors in Introductory Computing Student Drawings}


\author{Annapurna Vadaparty}
\orcid{0009-0002-4370-764X}
\affiliation{%
  \institution{University of California, San Diego}
  \city{La Jolla}
  \state{California}
  \country{USA}}
\email{avadaparty@ucsd.edu}

\author{Devamardeep Hayatpur}
\orcid{0000-0001-5984-9752}
\affiliation{%
  \institution{University of California, San Diego}
  \city{La Jolla}
  \state{California}
  \country{USA}
}
\email{dshayatpur@ucsd.edu}

\author{Adalbert Gerald Soosai Raj}
\email{asoosairaj@ucsd.edu}
\orcid{0000-0002-6848-2208}
\affiliation{%
  \institution{University of California, San Diego}
  \city{La Jolla}
  \state{California}
  \country{USA}
}

\author{Leo Porter}
\orcid{0000-0003-1435-8401}
\affiliation{%
 \institution{University of California, San Diego}
 \city{La Jolla}
 \state{California}
 \country{USA}}
\email{leporter@ucsd.edu}

\author{Daniel Zingaro}
\orcid{0000-0002-1568-4826}
\affiliation{%
 \institution{University of Toronto Mississauga}
 \city{Toronto}
 \state{Ontario}
 \country{Canada}}
 \email{daniel.zingaro@utoronto.ca}

\renewcommand{\shortauthors}{Vadaparty et al.}

\begin{abstract}
\textbf{Background and Context.} Problem decomposition is a core concern of computing education. It has also become increasingly relevant: in response to GenAI, many CS1 educators are advocating for shifting instructional emphasis away from code writing and towards decomposition and higher-level planning. Currently, there is a lack of knowledge in how novices do decomposition in large, multifunction tasks. 

\noindent\textbf{Objectives.} In this study, we describe how students represent solutions to a decomposition task, and characterize common issues that arise in those representations.

\noindent\textbf{Method.} In a 50-minute lab, students were given a description of a word game and asked to draw (with pencil and paper) a decomposition diagram for a program that would implement this game. We performed an inductive thematic analysis with negotiated agreement on 55 of the diagrams, coding salient elements (e.g. functions and the relationships between them) and issues that arose. 

\noindent\textbf{Findings.} Students used multiple representational strategies, including hierarchical function calls and sequencing (order of execution). We identified issues in notation (including use of differing, incompatible notations within the same diagram), order of execution, abstraction and reuse, encapsulation, clarity, and problem-specific misunderstandings. 

%\paragraph{Implications.} 
\noindent\textbf{Implications. }These findings suggest that novice decomposition is shaped by multiple underlying models of program behavior, with tensions between structural and sequence-focused reasoning. We discuss implications for decomposition instruction and future work, including clarifying representational constraints and plan tracing as simulation.
\end{abstract}



\newcommand{\todo}[1]{%
  \textcolor{gray}{%
    $\langle$TODO\ifstrempty{#1}{}{:\ #1}$\rangle$%
  }%
}
\newcommand{\dev}[1]{}
\newcommand{\av}[1]{}
\definecolor{BEIGE_BG}{RGB}{254, 254, 245}

\tcbset{on line, 
    boxsep=3pt, left=4pt,right=4pt,top=4pt,bottom=4pt,
    colframe=black!100,
    colback=BEIGE_BG,
    boxrule=0.5pt,
    enhanced,
    toptitle=1pt,
    bottomtitle=1pt,
    arc=0.5mm,
    drop small lifted shadow
}

\newenvironment{card}
{ 
   % \begin{minipage}[t]{0.74\textwidth}
    \begin{tcolorbox}[]
    % \sffamily
    \footnotesize
}
{
 \end{tcolorbox}
 % \end{minipage}
}

% Fancy section label
\makeatletter
\renewcommand{\sectionautorefname}{Section \S\@gobble}
\makeatother
\let\subsectionautorefname\sectionautorefname
\let\subsubsectionautorefname\sectionautorefname


\received{27 February 2026}
\received[revised]{1 May 2026}
\received[accepted]{12 May 2026}

\maketitle

\vspace{-0.5em}
\section{Introduction}
%P1: Decomp is foundational in CS ed, including in the GenAI era
Problem decomposition, the process of breaking complex problems into smaller, more manageable parts, is recognized as a foundational metacognitive skill in computing~\cite{barr2011bringing, wing2006computational}. Computing education research has advocated for explicit teaching of decomposition in introductory courses, emphasizing its role in abstraction, modular reasoning, and effective program design~\cite{felleisen2018design, haldeman2026systematically, muller2007pattern}. In light of the recent rise of GenAI tools, many educators have advocated for shifting core competencies of CS1 towards higher-level planning and decomposition~\cite{porter2024learn, winters2026cs}. An understanding of how novices approach decomposition may inform pedagogies for these shifting competencies. %Understanding how novices approach decomposition remains an open problem in computing education.

%P2: Prior work tells us about planning for small programs but not large ones
Prior work has examined how novice programmers plan solutions for small, code-based representations and has shown that novices rely on familiar schemas or patterns in a solution~\cite{soloway1986planning, spohrer_soloway_1986novice, rist1989schema, pennington1987stimulus}. They often stick too closely to these familiar schemas, have difficulty identifying boundaries for components of their plans, and underspecify their plans. These works focus on relatively small programming problems (e.g. searching a list or computing a maximum) where planning elements correspond to localized, well-defined code elements. We do not yet know whether the same issues arise when novices are faced with planning larger programs without pre-specified planning elements. Investigating how novices plan in complex, unstructured contexts is critical for understanding how planning and decomposition skills transfer beyond small, clearly constrained problems.

%Introductory programming increasingly asks students to reason about larger programs in order to better prepare them for future coursework or roles, yet we do not know how novices decompose problems at a higher level before implementation. % This is especially relevant when students must decide for themselves what the parts of a program should be and how those parts relate.
%Soloway, Rist, and others theorize

%P3: Prior work is also more structured but we are doing less-structured, student-generated artifacts, which helps us see more authentically what they're doing
To better characterize students' plans in less constrained contexts, we examined student \textit{drawings} of decomposition diagrams. In prior work, students decomposed problems into pre-defined code blocks (or templates) which may provide structure for a students' own model of how a problem ought to be broken down.
In contrast, a drawing provides a more authentic view of how students conceptualize plan elements (e.g. functions) and relationships because it allows greater freedom in representation.% , revealing both the structures they consider meaningful and the representational choices they make.

%P4: goal, RQs, methods
We explore how students conceptualize the decomposed parts of a program, what kinds of relationships they express between those parts, and the issues that emerge in these decompositions, asking the following research questions:
\begin{enumerate}
    \item[\textbf{RQ1:}] How do students represent and communicate the elements of their decomposition?
    \item[\textbf{RQ2:}] What issues do we observe in students' solutions for a decomposition task? 
\end{enumerate}
To investigate these questions, we conducted a study in which students were asked to create a decomposition diagram for a multi-function programming task, producing diagrams that included nodes (such as boxes representing functions), textual descriptions, and relationships between components. We then performed a qualitative analysis of these drawings to characterize both representational patterns and issues in students’ planning.

%P5: findings and implications
Our study reveals substantial variation in students' representation of decomposition, with two styles emerging: hierarchical function-call structures emphasizing abstraction, and sequencing-based representations that foreground execution and control flow. Students also used many new and ad-hoc notations outside of those used in instruction. For example, they introduced nodes representing conditional statements
and other annotations, suggesting aspects
of programs that are salient in their conceptual models beyond what their instructional materials included. They also sometimes (incorrectly) combine hierarchical and sequencing notations, indicating tensions between these two ways of reasoning. Other frequent issues included difficulties with encapsulation and repeated functionality. Our findings reveal a need for clearer instructional scaffolding around decomposition representations and highlight how understanding students’ planning artifacts can inform teaching practices that better support abstraction and program reasoning.

\vspace{-0.5em}
\section{Background and Related Works}\label{sec:lit_theory_related_works}

We ground our work in existing theories of how learners plan and decompose programs, as well as how they represent these plans through drawings. ~\autoref{sec:lit_planning} describes how learners plan and decompose problems and the types of errors they make when planning. We then turn to notional machines as a theoretical underpinning for our interpretation of drawings as representations of students' mental models in ~\autoref{sec:lit_notionalmachines}. These studies provide support for analyzing people's drawings to deepen our understanding of their mental models. 

%Some sort of sentence here.

%We differentiate from prior work along X dims:
%1. Study freeform drawings, rather than code.
%2. Study a sufficiently complex problem that requires reaching for some form of abstraction.
%3. Why are we interested in planning/decomp, and why are we interested in notional machines? Those are theoretical foundations and we are on the applied side--we use those theoretical lenses to explain why we're studying this 
\vspace{-0.5em}
\subsection{Planning and Decomposition}\label{sec:lit_planning}

%P1: Planning is decomp + composition
% There are many veins of related literature that help inform the processes in \textit{planning}. 
\citeauthor{fisler_shriram_2016modernizing} and ~\citeauthor{rivera_shriram2024observations_sketches} describe \textit{planning} as a two part process: breaking a complex problem down into simpler parts (problem decomposition), and then using these parts to create a solution (plan composition)~\cite{fisler_shriram_2016modernizing, rivera_shriram2024observations_sketches}. 
%P2: Decomp best practices
\textit{Problem decomposition} is the process of breaking a problem down into simpler constituent parts, and several conventions and best practices for decomposition in computing have been identified to enhance components' reusability and ease of use. Parnas argues that effective decomposition requires information hiding, where modules encapsulate design decisions likely to change and expose only necessary functionality~\cite{parnas1972criteria}. Parnas along with Stevens et al. state that decomposition should promote cohesion within modules (to support understandability of a module's purpose and ease of ability to maintain) and reduce dependencies and coupling between them (to support module reuse)~\cite{parnas1972criteria, stevens1974structureddesign}. When we use the term ``decomposition'' in our study context, we refer to the overall process of breaking a problem down into constituent parts and relating those parts together, encompassing aspects of both composition and decomposition in the planning literature.
%
%P3: How people do decomposition and planning
%
There are various explanations for how people perform problem decomposition. One strategy is \textit{subgoal decomposition}, which requires breaking down a problem as a sequence of subgoals in a particular order~\cite{corbett1995knowledge, catrambone1998subgoal}.
 An alternate strategy is through a \textit{means–ends analysis}, in which problem solvers iteratively reduce differences between current and goal states by generating intermediate subgoals~\cite{newell1972human}. 

\subsubsection{Planning and Decomposition Practices in Novices}\label{sec:lit_planning_novices}
\vspace{-0.5em}
%P1: novices focus on local goals & immediate steps 
Novices face challenges in planning and decomposition that stem from local rather than global goals. Soloway’s work on program plans suggests that programming expertise involves recognizing and applying reusable higher-level schemas, whereas novices often instead reach for immediate implementation~\cite{soloway1986planning}. Linn and Dalbey characterize novice programming knowledge as fragmented, with learners frequently constructing solutions incrementally rather than from a coherent global plan~\cite{linn1985cognitive}. Taken together with the subgoal decomposition strategy (described earlier),
this could be explained as novices using a process that resembles a greedy algorithm, where each next subgoal is selected based on immediate salience rather than a global strategy. 

%P2: People use bottom up when they're unfamiliar, and higher level representations when they're familiar (focal expansion when they know some parts but not others) 
Novices tend to use low-level representations and bottom-up planning strategies for portions of code they are unfamiliar with, but work top-down from known schemas when they recognize a pattern. Song and Becker found that novices approach problem solving using concrete execution details rather than abstract functional organization, and construct plans bottom-up once they have identified concrete components~\cite{song2014expert, song2016problem}. These results align with Pennington's earlier finding that novices prefer lower-level procedural representations ~\cite{pennington1987stimulus}, and other findings that novices focus on syntax-level implementation~\cite{linn1985cognitive, robins2003planning}. Novices' use of bottom-up approaches is also reiterated in Rist’s focal-expansion model, which describes two problem solving approaches: top down, in which people retrieve a known solution to a similar problem to address the current one, or bottom-up by creating a solution when no similar example exists by starting from a small ``focal'' code fragment ~\cite{rist1989schema, rist1991knowledge}. When a known solution only partially applies to a harder problem, students may begin with top-down retrieval and then shift into bottom-up creation to adapt it. As programmers gain experience, they rely increasingly on plan retrieval rather than focal expansion ~\cite{rist1991knowledge}.

%P4: Issues
Spohrer \& Soloway synthesize a variety of novice programming errors including ones related to planning~\cite{spohrer_soloway_1986novice}. While their findings were from a study of concrete code, some of the errors remain relevant in more freeform contexts like ours. For example, the ``summarization problem'' occurs when implications of earlier parts of plans are overlooked and have unintended consequences on later plan components, and ``the boundary problem'' describes difficulties novices have with boundary points in plans. Similar boundary issues are found in more recent work such as Muller et al.'s study that identifies erroneous ``blending of components'', and they provide an example of incorrectly using a common variable to iterate through both an inner and outer loop~\cite{muller2007pattern}. 
%DZ suggesting this cut (it's... I can see peripherally why we'd mention it, but it isn't specific to decomp, and we don't return to it)
%Outside of planning errors, Spohrer and Soloway also describe the ``human interpreter problem,'' in which novices assume that the computer shares an understanding of their intention of what a program should do or how a construct should be interpreted; this is reminiscent of Roy Pea's ``superbug'', described as ``the default strategy that there is a hidden mind somewhere in the programming language that has intelligent interpretive powers''~\cite{pea1986superbug}. 

We seek to understand whether novice tendencies found in these prior studies on concrete code (locally driven subgoals, bottom-up reasoning, and boundary-related errors) also surface in the freeform medium of drawings. % This is in contrast to code-based artifacts used by prior studies.
%DZ cut here because it felt like a stronger conclusion to this section without this next sentence
%The task we select for our study (~\autoref{sec:methods_task}) is sufficiently complex that we suspect students will not be able to visualize it at the level of code when asked to decompose it. %We wish to see whether and how the issues present in their solutions differ from prior work in code-based problems. 
\vspace{-0.5em}
\subsubsection{Planning and Decomposition Teaching Practices} \label{sec:lit_planning_teaching}
There are various ways to teach decomposition as a computational skill. De Raadt and Toleman argue that teaching and assessing programming plans directly can help learners develop awareness of planning processes and improve their ability to structure solutions before implementation~\cite{deraadt2009teaching}, a finding shared in Loksa et al.'s study of students given explicit guidance when working through problem-solving steps~\cite{loksa2016programming}. Rich et al. introduce a variety of ways in which a problem can be broken down; one such way is into different components of a solution, and another is in the assumptions or requirements set forth in the problem~\cite{rich2019framework}. Pattern-oriented instruction described by Muller et al. further provides learners with reusable organizational structures to help them identify patterns in new problems they encounter~\cite{muller2007pattern}, which Haldeman et al. reference in their framework for strategies of teaching decomposition~\cite{haldeman2026systematically}. Finally, instruction on plan composition emphasizes recombining smaller pieces into coherent solutions, helping learners connect decomposition with higher-level program organization~\cite{rivera2022plan}. 
%These works emphasize the importance of explicit instruction and scaffolding to help students understand how to break down and recompose problems. 
These works motivate our focus on students' free-form representations of decompositions: We can identify gaps that may exist between intended outcomes of current instructional practices and the decompositions that students create. %\newline AV: add newline back in for Camera Ready

% In our efforts to understand students' decomposition practices, we seek prior work that may provide a lens for understanding the representations they create. We next discuss the relevance of notional machines in interpreting planning artifacts. 

Most prior studies of novice planning in computing involve small programs and concrete, code-based environments. This leaves open the question of how novices approach larger programs and less constrained contexts, which motivates both dimensions that differentiate our study from prior work: we analyze freeform drawings, and study a sufficiently complex problem that requires students to reach for some form of abstraction (described in \autoref{sec:methods_task}). Both of these study design choices allow us to explore how students represent plans when they cannot rely on the structure and constraints of code. To ground our interpretation of student drawings, we next turn to notional machines.
\vspace{-0.5em}
\subsection{Notional Machines}
\label{sec:lit_notionalmachines}
\citet{duboulay1981blackbox} coined the term \textit{notional machine} to refer to ``the idealised model of the computer implied by the constructs of the programming language.'' \citet{sorva2013} shares and builds upon this definition by adding that notional machines model runtime behavior, provide a language-specific perspective for explaining program semantics, and accurately reflect how programs execute in practice. The computing education community has used various meanings of notional machines, with the 2020 ITiCSE Working Group arriving at the following definition: ``A notional machine (NM) is a pedagogic device to assist the understanding of some aspect of programs or programming''~\cite{fincherITiCSE2020notional}. Although we do leverage this pedagogical lens in our current work, we more closely follow duBoulay and Sorva's characterizations of notional machines without the requirement for them to be explicitly created with a pedagogical intention.

It is helpful to distinguish between a student's mental model and the notional machine. As described by \citet{cunningham2017studentsketches} in the context of tracing, ``When a student writes code with an expected outcome or predicts the result of running a piece of code, the student runs that code through their mental model of the notional machine.'' More generally, a student has a mental model \textit{of a} notional machine--that is, they have a mental model (which may or may not be correct) of an abstract/idealised model of a computational process (the notional machine). Teaching, then, aims to align students' mental models with an accurate notional machine. This perspective informs our use of student-generated drawings as artifacts that externalize their model of an underlying notional machine.
\vspace{-0.5em}
\subsubsection{Procedural and Object-Oriented Notional Machines}\label{sec:lit_notionalmachines_proceduralOO} A variety of notional machines have been identified. \citet{sorva2013} identify two broad classes: imperative/procedural notional machines (the focus of this work), and high-level object-oriented notional machines ~\cite{sorva2013}. The imperative notional machines involve variables and a program counter for order of execution~\cite{sajaniemi2008procedures}, whereas the object-oriented notional machines have execution control occurring ``inside the objects''~\cite{bergin2000procedural}---order of execution is still necessary to keep track of in this notional machine, but it happens within elements (objects) instead of as relationships between them or a visible global sequence. Sorva conceptualizes the higher level object-oriented notional machine as a composition of two notional machines: ``One can be seen as an object-enabled extension of a procedural notional machine...and another describes message-passing between interacting objects,'' and notes that using both of these notional machines simultaneously is ``more challenging to teach successfully.'' Sorva also notes that students may hold multiple, possibly contradictory mental models of a notional machine.
%, and our current work may shed light on whether incorrect merging of two notional machines may be one such contradiction.
Later, we provide concrete examples of incorrect and contradictory blends of notional machines that appear in our analysis of student drawings.

\vspace{-0.5em}
\subsubsection{Drawings can externalize mental models of notional machines.}
Drawing and sketching can serve as a useful method for eliciting students' mental models \cite{jones2011mentalmodels, williams2003dataabstraction, chi2001learning}. Drawings force the resolution of certain ambiguities in mental models before they can be drawn~\cite{dix2011externalisation}. Internal mental models can remain vague or underspecified (e.g. a learner may know a loop is present without deciding where it should be), but producing a drawing requires committing to  those details and can therefore surface knowledge gaps or misconceptions. Mayer's framework on multimedia learning suggests that students need to select, organize, and integrate information in their mental model in order to be able to represent it~\cite{mayer2002multimedia}. Relatedly, \citet{chi2001learning} find that drawing exposes a learner's evolving conceptual understanding as it becomes closer to what they are being taught. Tversky's seminal work on spatial representations~\cite{tversky2011visualizing} posits that people recruit spatial structure in order to convey conceptual structure, i.e. that the spatial relationships in one's drawing reflect one's underlying mental model. Others have also shown that sketches help ground abstract ideas through spatial metaphors~\cite{hendry2006sketching, blackwell2001cognitivefactors}. In computing education, sketching and diagram-based activities have been used to provide insight into how novices represent program structure and execution~\cite{cunningham2017studentsketches, whalley2007decoding}. These works ground our use of drawings as a way to understand student mental models of the solution they are designing. 

\vspace{-0.5em}
\section{Methods: Course Context and Task}\label{sec:methods_course_and_data}

\subsection{Course Context \& Decomposition Instruction}\label{sec:course_context_decomp_instruction}
\begin{wrapfigure}[9]{r}{0.4\textwidth}
  \centering
    \includegraphics[width=1\linewidth]{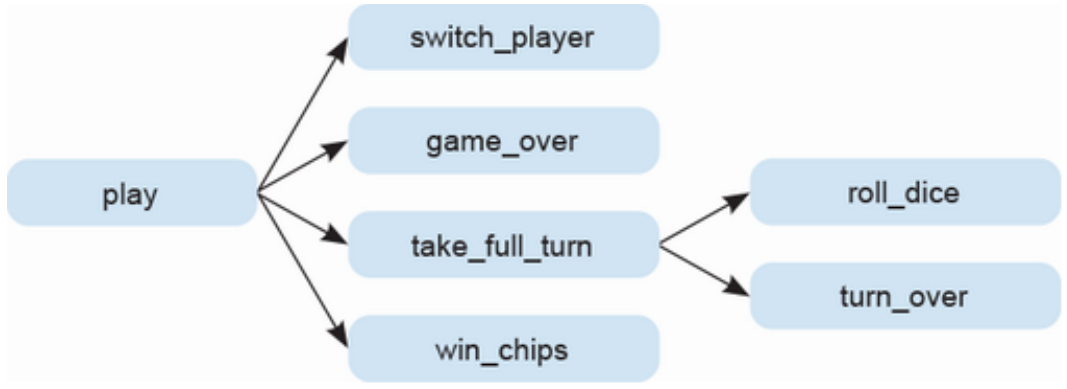}
    \caption{An example of a hierarchical function call decomposition diagram from the course textbook for a two-player dice game (\citet{porter2024learn}). }
    \label{fig:textbook_game_decomp} 
\end{wrapfigure}
Our study draws from students in a large introductory CS1 course in Python that had been recently redesigned to include shifting core competencies in light of GenAI. The course taught debugging, testing, working with GenAI tools, and problem decomposition, as well as traditional CS1 skills such as tracing code and manually writing some code from scratch. We used the textbook \textit{Learn AI-Assisted Python Programming} by \citet{porter2024learn}. Problem decomposition was covered through readings, multiple lectures, homeworks, and course projects.  In the readings, students learned about the need to break apart large tasks into smaller functions. These readings emphasized best practices such as creating functions with clear responsibilities, avoiding monolithic ``do-everything'' functions, and favoring reusable helpers over duplicated logic in order to facilitate testing and isolate where in the code bugs arise. Readings mentioned the importance of functions having the following properties: a clearly defined purpose, not being too long (roughly 12-20 lines of code), general and reusable, and clear inputs and outputs. These chapters showed multiple concrete examples of large tasks being broken into subfunctions with corresponding diagrams, and Python programs built using those diagrams as a guide. Similarly, in lecture, the problem decomposition instruction included diagrams that were translated into fully coded solutions where the instructor referred back to the diagram to guide the coding of each function.

% \paragraph{\textcolor{red}{A modification of Leo's version Feb 25.}} 
The primary diagrams shown when teaching problem decomposition were hierarchical function call diagrams modeled after call graphs in software engineering in which arrows from a caller function denote it calling a helper function~\cite{grove1997call_hierarchical, latoza2011visualizing_hierarchical}, an example of which is shown in \autoref{fig:textbook_game_decomp} from the course textbook. In this example, ``play'' calls the four child functions, one of which is ``take\_full\_turn'', and this function calls ``roll\_dice'' and ``turn\_over''. After the course offering, the first author reviewed the course content, including material developed during instruction by teaching staff (e.g. for discussion sections), and found diverging notations. These included data flow (where arrows represent data streams) and execution order (where arrows represent one function being executed after another). These alternate depictions often resembled control flow diagrams, another visualization used in software engineering~\cite{zhu1995axiomatic_controlflow, white1987software_controlflow}. Given the scale of the course (13 teaching staff members and 181 students), this diversity in notation is unsurprising. Empirical studies, even among software developers, show that diagrams are often context-dependent rather than adhering to a standard notation~\cite{cherubini_Amy_Ko2007whiteboard, storrle2013towards}. We suspect that a lack of a strict standardized notation contributed to the inconsistent representations in student diagrams. Therefore, although hierarchical diagrams were the primary way that students were taught to perform problem decomposition, the presence of alternate notations led to students seeing multiple diagramming approaches. %However, we note that our large-course conditions with several teaching staff members mirror those of many CS1 offerings, and this lack of standardization is common in early iterations of courses that redesign components. 

Students had multiple-choice homework and exam questions in which they were asked to choose among several options for diagrams with the hierarchical function notation. The homeworks included relatively complex problem descriptions (similar in difficulty to our current task described in the next section) and asked students a series of questions. These began with comprehension check questions to ensure that students were understanding the problem statement itself and the intended behavior of a solution, followed by the question in which they were asked to select the diagram that best represents effective decomposition. %AV: should we add a sentence here explaining why we assessed with hierarchical function calls? My above paragraph was intentionally neutral about one being better than another so as not to introduce questions about why we did both if one is better. 
Additionally, students had three open-ended programming projects throughout the course where they submitted their code, a video code walkthrough, and a decomposition diagram. In their diagrams, students were required to include the \textit{description}, \textit{inputs}, and \textit{outputs} for each function. Students received a score on their diagrams based on a rubric that had point deductions for errors such as not providing function inputs or outputs, not providing descriptions of functions, or not including all functions that were found in their code. As such, students had previously drawn decomposition diagrams and received rubric grading feedback on those diagrams prior to our study.

\paragraph{Lab and Data Collection} Students had 50 minute weekly lab sessions, and their task for the lab in the ninth week of our 10-week course was to draw a decomposition diagram for Evil Word Guesser (described below). They received full marks for any submission (graded for completion). The writeup describing the task had a description of the game, an example showing a guess and explaining the selection of smaller and smaller lists of words (referred to as ``word families''), and the instructions: ``Please draw a diagram (like the decomposition diagrams you’ve created for your projects) showing which functions you would break the game into, including the interactions between the player and the evil computer. For each function, write a simple description of what the function should do (you can write this anywhere as long as it’s clear–it can be next to the function in your diagram, or you can write all your function descriptions off to the side).'' Because these instructions only explicitly mentioned including a description for functions, we did not expect students to additionally include function inputs and outputs. Students drew their solutions on an 8.5"x11" sheet of paper, which were then scanned and anonymized. In total, we had 133 student solutions. Data collection and analysis for the study were approved by our Institutional Review Board. 

\subsection{Evil Word Guesser Task}\label{sec:methods_task}
\begin{figure}[h]
\centering
\begin{minipage}[c]{0.4\textwidth}
\begin{card}
% \textit{\textbf{\normalsize{Evil Word Guesser:}}} \\

\sffamily{
\textit{\textbf{Evil Word Guesser:}}
The computer never actually picks a specific word! Rather, it starts with the set of all possible words of a given length and always chooses the pattern that leaves the most options. This makes it much harder to guess the correct word! \\

For example, if you guess ``e'', the computer groups all possible words into word families based on where (or if) ``e'' appears in them. One family would be all words that follow the pattern ``\_ \_ e'', and another family would follow the pattern ``\_ \_ \_'', and another would be ``e \_ \_''. The words ``ate'' and ``the'' would be in the word 
%DZ: added a space before e. Is this correct?
family ``\_ \_ e'', and ``ear'' and ``elk'' would be in the word family ``e \_ \_''. The game then picks the largest family--the one that leaves it the most options--to keep the game going as long as possible!
}
\end{card}
\caption{An excerpt from our Evil Word Guesser task description given to students.}
\label{fig:problem-description}
\end{minipage}\hfill
\begin{minipage}[c]{0.55\textwidth}
    \centering
    \includegraphics[width=1\linewidth]{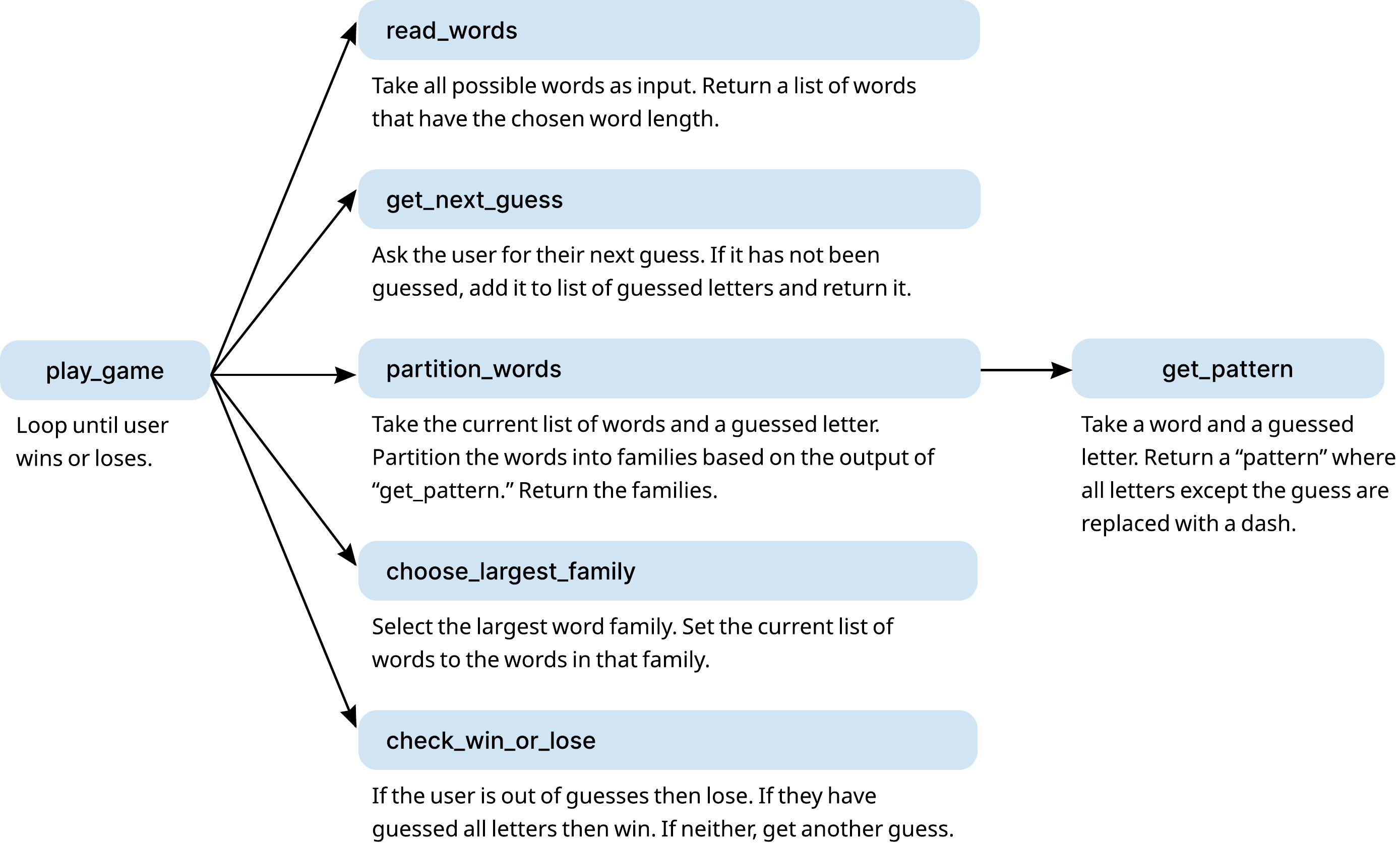}
    \caption{A sample hierarchical diagram solution.}
    \label{fig:sample_hierarchical_diagram}
\end{minipage}
\end{figure}

We asked students in the introductory computing class to draw a function decomposition diagram as a lab activity as described in the previous section. Specifically, they were asked to implement a variant of `Hangman,' a popular word game where players guess letters one at a time to uncover a hidden word, losing a chance for each incorrect guess until the word is revealed or their allowed mistakes run out. In the variant, the game does not store a single correct word as the solution, but instead maintains and updates the maximum-size list of words that are all compatible with the user's guess (see \autoref{fig:problem-description}).\footnote{The Evil Word Guesser (Evil Hangman) task: http://nifty.stanford.edu/2011/schwarz-evil-hangman/} A sample solution to this task that uses the hierarchical notation is shown in ~\autoref{fig:sample_hierarchical_diagram}. 

This task is complex enough that a student would be unlikely to visualize the code and work backwards to create a decomposition diagram. %; in other words, some functions would need to stay as ``black boxes'' in order for them to finish the task in the allotted time.
 We are therefore able to see how students decompose a problem through the diagram at the level of abstraction they see fit. This is in contrast to prior studies of planning and decomposition in CS1 that use simpler exercises 
%DZ could we cite this (rainfall Soloway paper) using something we already cited?
(like the Rainfall problem studied by Soloway~\cite{soloway1986planning}). %, where the student solution is code %(like the higher-order functions referenced in ~\autoref{sec:lit_planning}).
Additionally, a correct solution must include an interaction loop, which allows us to examine how students represent control-flow constructs like repetition.

\subsection{Thematic Analysis of Student Diagrams}

\paragraph{Thematic Analysis of Student Diagrams}\label{sec:thematic_analysis}
% As described in \autoref{sec:lit_theory_related_works}, we find ourselves working with a dataset that is more freeform and unstructured than those in many prior works. 
The first two authors conducted a 
codebook thematic analysis of student diagrams. We began by creating a random ordering of diagrams for analysis, and then sampled diagrams from this ordering. For each diagram, the lead author transcribed the features we deemed necessary in our analysis: ~\textit{nodes} (e.g. boxes), and ~\textit{relations} between nodes (e.g. arrows). To answer RQ1, we coded the \textit{type} and \textit{purpose} of each node (e.g. a \textit{function} that \textit{generates word families}) and relation (e.g. an arrow that indicates a \textit{hierarchical function call}). 

To answer RQ2, we then coded each diagram for the issues present. We noted missing functionality (e.g. not having a game loop) along with issues in decomposition. To identify missing functionality, two experienced instructors on our research team, both of whom redesigned the course and one of whom was the course instructor, along with the lead author, discussed and agreed on required functionality in a correct diagram (e.g. existence of a game loop). These functionalities are listed in \autoref{sec:appendix_elements_and_purposes}. 
% This informed our decisions about whether or not a diagram was missing functionality, described further in ~\autoref{sec:results_errors}. 
The first two authors, drawing from their prior experience as CS1 teaching staff, also coded other potential issues in student solutions (e.g. use of inconsistent notations and poor decomposition). The focus of our study is on decomposition and planning, and to this end, we did not include certain low-level mistakes if they did not lead to confusion or inconsistencies in the diagram as a whole. We therefore ignored issues such as misuse of game terminology or syntax errors for students who included code/pseudocode as long as the solution was understandable and could lead to a plausible correctly-functioning program.

We (the two coders) met over the duration of 11 weeks, coding the nodes, relations, and mistakes in each diagram individually and meeting to discuss codes each week. We began by independently coding the first 15 diagrams,  and then discussed these diagrams and created an initial codebook, following a bottom-up inductive thematic analysis process (sometimes referred to as ``codebook thematic analysis'') ~\cite{braun2019reflecting_codebookTA, padiyath2026reflecting}. After this initial batch, we continued to independently code and then met to discuss our codes for each diagram, iteratively refining our codebook. We resolved disagreements using negotiated agreement~\cite{campbell2013coding, hammer2014confusing}. We stopped this process when we were not able to add new codes or make alterations (i.e. reached saturation) and, crucially, our belief that we had substantive answers to our research questions~\cite{braun2021saturate, small2009many}.
As we iterated on the codebook, we updated previously coded diagrams, deductively coding for new issues. In total, we coded 55 diagrams. 

%https://andrewhead.info/assets/pdf/augmented-formulas.pdf maybe look at this for methods 

\subsection{Limitations}
Before proceeding to our results, we acknowledge the following limitations of our methodology. 

%\paragraph{Generalizability concerns from our CS1 course that included decomposition instruction}Students in our population might not be generalizable to all novices because they came from a course  explicitly taught decomposition.  %nonetheless, the patterns and mistakes they made indicate that they often diverged significantly from instruction, suggesting that novices may make these mistakes independent of the specific nature of their instruction. %\textcolor{red}{is this too strong a claim? }
\paragraph{Single institution and course with unique decomposition instruction} Our study participants were students at the end of a single CS1 course at a U.S. university. Additional research would be needed to examine whether and how these findings generalize across CS1 courses at other institutions. In addition to institutional, geographical, and policy differences, different CS1 courses may vary in how extensively and in what ways decomposition is taught. Instructors using different representations or pedagogy may find different results in student diagrams. 

\paragraph{Varied representations in instructional diagrams} In addition to the core instruction that represented decomposition with hierarchical function call notation, students were exposed to notations that more closely resembled control flow or data flow diagrams. This may have led to some of the variation seen in the results, making it unclear whether students would have made different choices if the instruction had adhered to a standard set of design decisions. While this limits the claims we can make in our work (internal validity), we believe this reflects the circumstances many large-course teaching staff may face, especially during curricular redesigns (perhaps offering external validity). 

\paragraph{Sampling} From our 181 student course, we had diagram submissions from 133 students. Because labs were a small part of the course grade, many students may have chosen not to attend, and our sample only reflects students present. 

\paragraph{Grading and diagram usage} The completion grading of this lab assignment may have led to students being less careful and intentional in their drawings than they may have been if we had graded for correctness, which may have led to over-representation of errors. Additionally, students were never asked to use these diagrams for implementation; knowing that they would never need to create and run a program based on their design may have decreased their motivation for creating a correct diagram, which may similarly have led to over-representation of errors. However, we maintain that there is value to planning-only exercises such as diagramming to encourage finding such errors before the implementation phase. Such planning-only exercises can help prepare students for smoother implementation phases in their own work, and also to equip them for contexts in which they may only be responsible for creating a working plan which they then give to another teammate or agent to implement.

\paragraph{Priming in bottom up coding} It is possible that conducting the expert consensus on necessary functionality could have affected our inductive coding of the diagrams; however, we proceeded with this approach because of the benefits we perceived in having concrete functionalities with which to ground our analysis. As discussed by Braun and Clarke~\cite{braun2006using_originalTA, braun2019reflecting_codebookTA}, there are a variety of stances regarding how much background knowledge should be considered when performing inductive coding, and we take the position that having a consensus about what functionalities were necessary in a working program was beneficial in having anchoring points in our analysis of such free-form, open ended data.

\section{Results: Representations and Decomposition Issues}\label{sec:results_solutions_representations}
We describe the representations students used (RQ1) along three dimensions:

%Here we describe the representations students used to show their solutions (RQ1) and the issues that arose (RQ2). We begin with an overview of these findings, and then walk through some illustrative example diagrams in~\autoref{sec:results_illustrative_examples}, pointing out attributes of student representations and some of the issues we find. Further detail describing these representations and issues are then in ~\autoref{sec:results_representations} and ~\autoref{sec:results_errors}, and are referenced throughout our illustrative examples. 

%\subsection{Overview}\label{sec:results_overview}

%We deconstruct the representation of student diagrams (RQ1) by identifying the following three attributes (detailed in ~\autoref{tab:node-relation-codebook}): 
\begin{itemize}

    \item \Node{Node types} (4 codes): \textit{What do the nodes represent?} Nodes were generally \textit{functions}, but were sometimes \textit{fragments} (i.e. small portions of functionality that were not enclosed in a function), \textit{data}, and \textit{annotations} (i.e. auxiliary details or instructions meant to describe other nodes or relations).
    \item \Purpose{Node purposes} (21 codes): \textit{What purpose do the nodes serve?} Each node usually held a game-specific purpose, like prompting the user for a guess, or checking if the win condition had been met.  
    \item \Relation{Relation types} (2 codes): \textit{How are the nodes connected to each other?} The lines or arrows sometimes represented a \textit{hierarchical function call}, where the relation denotes one function calling another function (a helper) to perform a subtask. Another type of relation was for \textit{sequencing}, where a relation between two nodes represented a chronological order of execution (i.e. one node is executed before the other). 

%DZ in Table 1, the node purposes come in the middle, not at the end like here. Should we move them to the end in Table 1 as well? %AV Good point--I've rearranged the bullets here (easier than rearranging the table and I don't think either order is particularly better or worse)
\end{itemize}

We also find \Error{Issues} (16 codes) in student solutions (RQ2). Some of these included issues with order of program execution, abstraction and reuse of functionality, encapsulation, clarity, and issues specific to the evil word guesser task.

\subsection{Illustrative Examples and Representations}\label{sec:results_illustrative_examples}

Prior to our full analysis, we start with three example diagrams to illustrate the breadth of  \Node{Nodes}, \Relation{Relations}, \Purpose{Purposes}, and \Error{Issues} we found.\footnote{{Alt text for figures is available \href{https://docs.google.com/document/d/e/2PACX-1vQIJ7csmTNZ6NlmjxJWolAOPQlWGXpL3l7VEbxN7mxZAstnBVexhHGaZhlvHafVQl8Ir6FOPIekjJb2/pub}{\underline{here}}. %We have also made efforts throughout the paper to have all key parts of the diagrams also described in captions. 
}}

\begin{figure}
    \centering
    \includegraphics[width=0.95\linewidth]{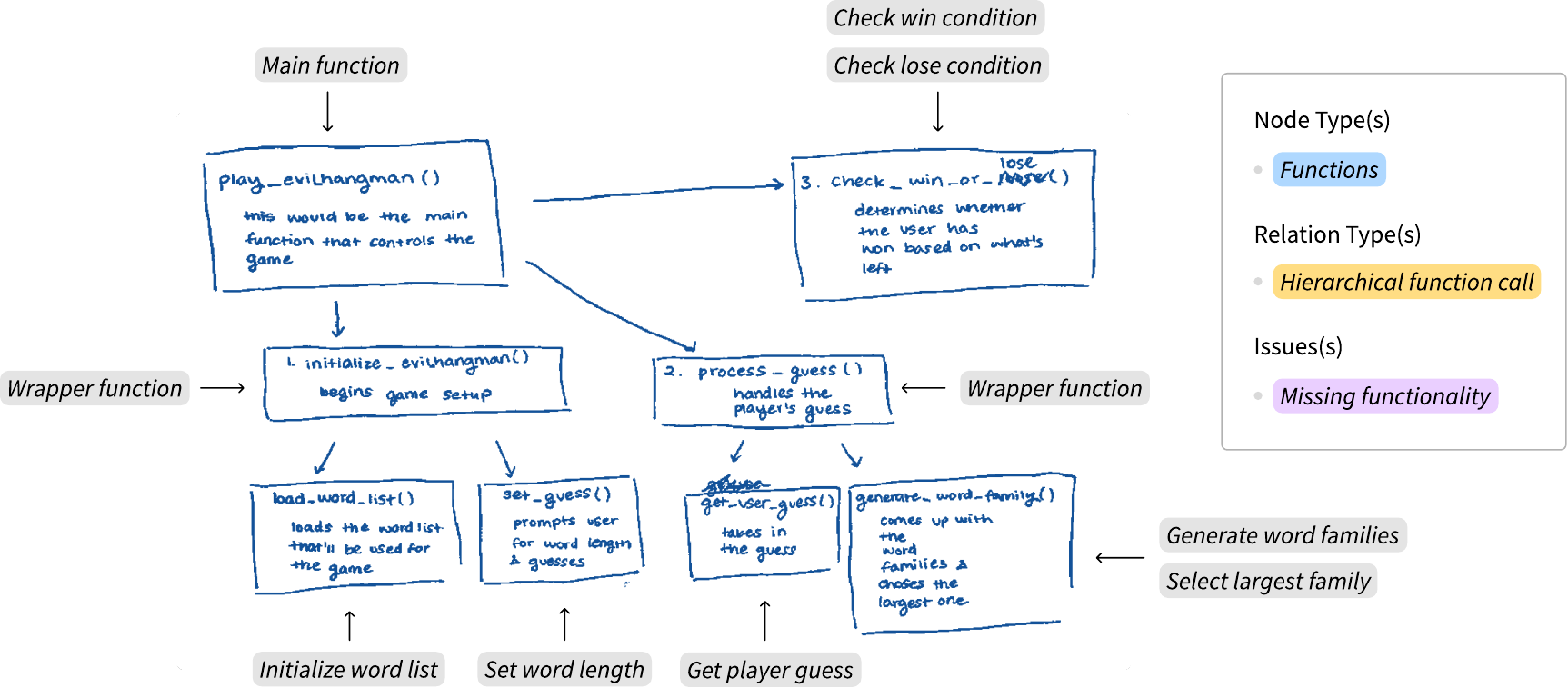}
    \caption{A diagram showing \nodecode{functions} making \relationcode{hierarchical function calls}, i.e. arrows represent a function calling a helper. Purposes of each node are labeled in grey throughout this illustrative example (and both others).}
    \label{fig:d7}
\end{figure}

\subsubsection{Illustrative Example 1}\label{sec:results_illustrative_1} We coded the diagram in ~\autoref{fig:d7} as follows.

\begin{itemize}
    \item \Node{Node Types} and \Purpose{Purposes}: The diagram decomposes the problem into \nodecode{functions}. The functions themselves hold various game-specific purposes. The ``play\_evilhangman'' in the top left of the diagram is a \purposecode{main function} that ``controls the game,'' which branches out to three helpers. The first of these (``initialize\_evilhangman'') is a \purposecode{wrapper function} around two helper functions. The first helper, ``load\_word\_list'', \purposecode{initializes} the word list while the other is used to 
%DZ all purposecodes should exactly match the codes in Table 1, right? So we should delete 'the'? Or, are we modifying slightly for cleaner sentences here. I think cleaner sentences is OK, so I vote to leave it alone, but I can't easily check/confirm the purposes that we have in the Figs and if they match the wording in Table 1. %AV We're modifying for cleaner sentences, so if that sounds ok I think we're good keeping it! Glad that's your vote too. 
\purposecode{set the word length}.\footnote{For brevity, we will not discuss all node purposes. Instead, the purposes are labeled in each diagram's figure and a definition for each code is provided in~\autoref{tab:node-relation-codebook}.}
    \item \Relation{Relation Types}: The nodes are connected to each other with arrows indicating \relationcode{hierarchical function calls}. For example, ``play\_evilhangman'' is a function that calls three helper functions. 
    \item \Error{Issues}: The diagram is largely correct but, like many other diagrams, 
    %does not explicitly describe repetition (like a loop), meaning that it has \errorcode{no/implicit interaction loop}. 
    % It alludes to multiple guesses happening, but never explicitly states this. 
    % Additionally, 
    it has \errorcode{missing functionality}. None of the nodes update the word pattern, and no loop is explicitly described. (Both of these are game elements required by instructors for a correct solution; see ~\autoref{sec:appendix_elements_and_purposes}). % none of the nodes have a purpose that \textit{update the pattern so far}, which is one of the required game elements specified by instructors as described in ~\autoref{sec:thematic_analysis}.
\end{itemize}

\begin{figure}
    \centering
    \includegraphics[width=0.95\linewidth]{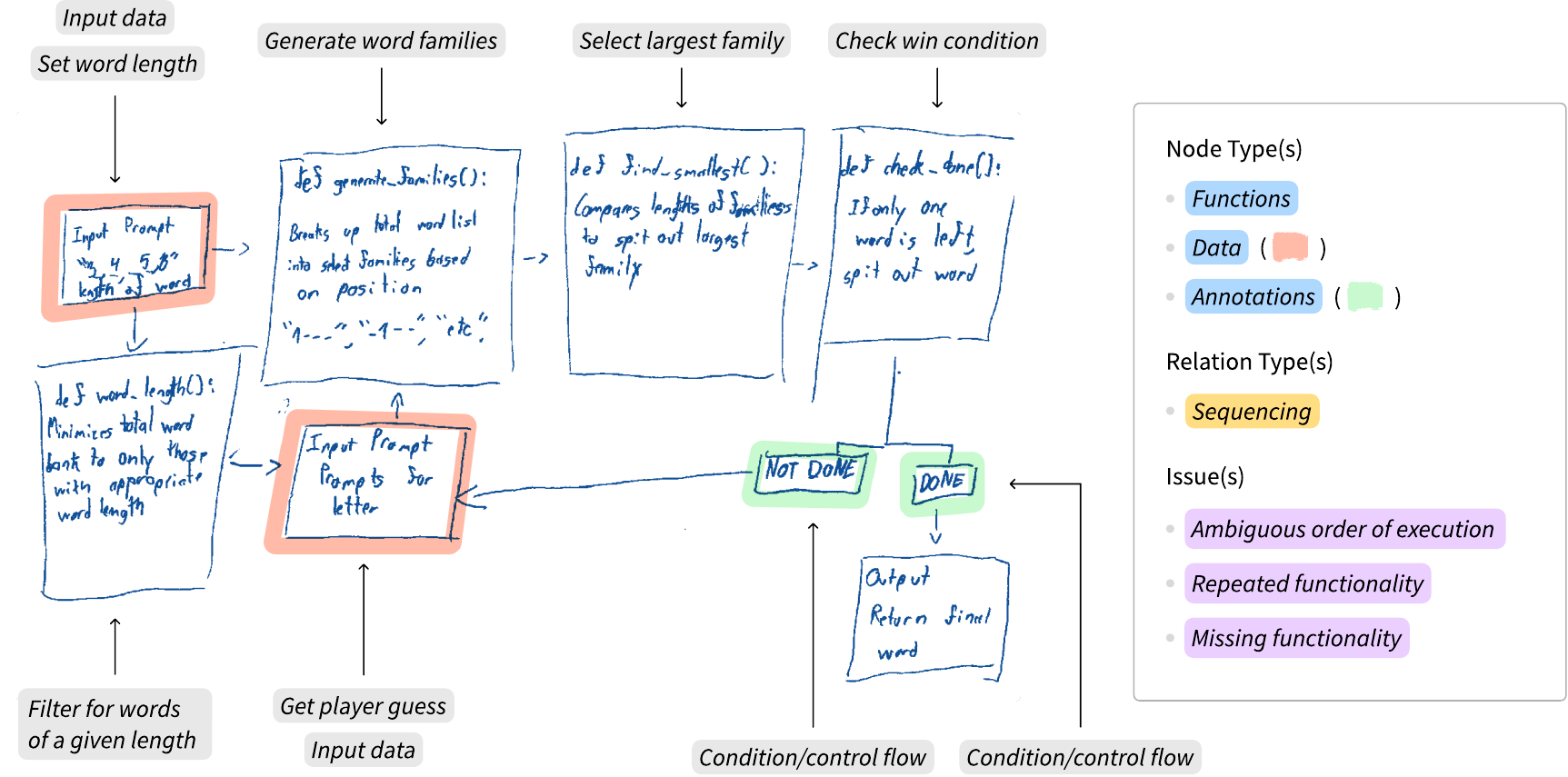}
    \caption{A diagram showing functions connected by arrows that indicate \relationcode{sequencing} of execution order. Most nodes are functions, with two \nodecode{data} nodes (highlighted in red) representing \purposecode{input data}. There are also two \nodecode{annotation} nodes highlighted in green that handle \purposecode{condition/control flow}, creating branching functionality. Less clearly specified are the relations that branch from the top-left input data node, which create an \errorcode{ambiguous order of execution} as it is unclear which of the two relations to follow after obtaining the input. There is also \errorcode{repeated functionality}, as both the top right ``check\_done()'' node and bottom right ``output'' node return the final word if the user has won.}
    \label{fig:d51}
\end{figure}

\subsubsection{Illustrative Example 2} \label{sec:results_illustrative_2}\autoref{fig:d51} illustrates further variety in representations: %The next illustrative example is the diagram in ~\autoref{fig:d51}. % We focus on types, purposes, and errors that were not covered in the previous example. 
\begin{itemize}
    %\item  \Node{Node Types} and \Relation{Relation Types} 
    \item  \Node{Node Types} and \Purpose{Purposes}: Like the previous example, several nodes are \nodecode{functions}. However, this diagram also includes two \nodecode{data} nodes, both of which are used to get \purposecode{input data}--one gets the length of the word and the other is for the player guess. 
    %There are also several \nodecode{functions}. 
    Moreover, the function ``check\_done()'' branches out to two relations that include \nodecode{annotations}. These ``NOT DONE'' and ``DONE'' nodes indicate \purposecode{condition/control flow}, i.e. whether or not downstream nodes should be visited. % The ``DONE'' node then points to the ``Output'' node, and ``NOT DONE'' points back to the ``Input Prompt'' node that gets the user guess.   

    \item \Relation{Relation Types}: Unlike the prior example, the relations in this diagram show procedural \relationcode{sequencing} between nodes. For example, after getting the length of the word from the top-left ``Input Prompt'' node, the word list is narrowed down to this length, and then the program prompts the user for a letter and begins gameplay. 
    % Relations show order of program execution. 
    To show a loop, a sequencing representation must include a ``back'' relation from later nodes to earlier nodes. Here, such a back relation is shown between the ``NOT DONE'' and ``Input Prompt'' nodes.
    \item \Error{Issues: } We find an \errorcode{ambiguous order of execution} if we follow the branching relations leaving the top left ``Input Prompt'' node. It is unclear which of the two branches, ``def word\_length'' or ``def generate\_families'', to take first. 
    %: is there a better word for back relation?
    
    %for the two ``Input Prompt'' nodes whose relations point to the ``def generate\_families()'' node. It is unclear which of the ``Input Prompt'' nodes is meant to be executed directly before ``generate\_families():''. 
\end{itemize}

\begin{figure}
    \centering
    \includegraphics[width=0.95\linewidth]{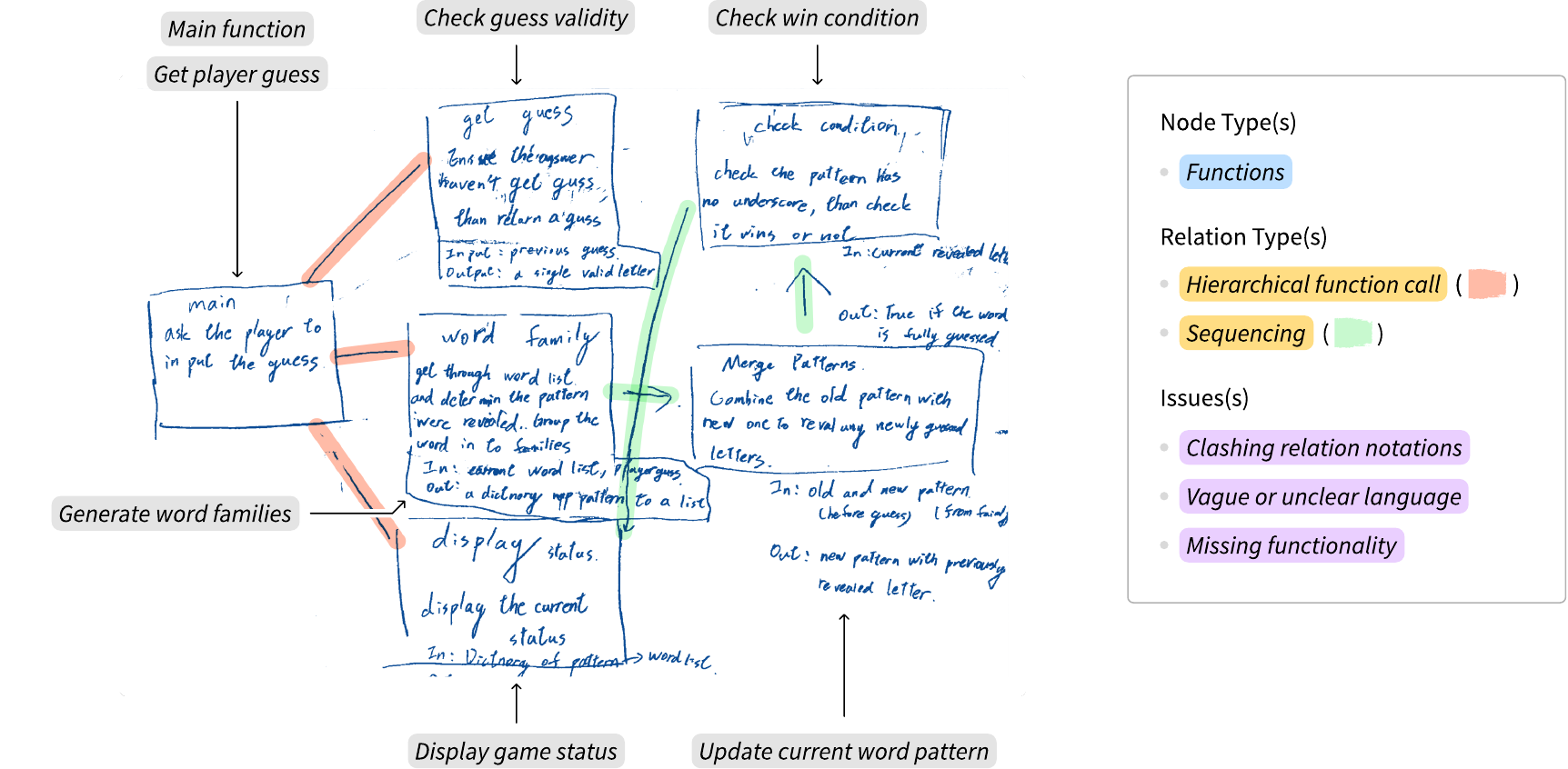}
    \caption{A diagram with \errorcode{clashing notations}. The three relations leaving the main function indicate hierarchical function calls (highlighted in red) and the remaining three relations indicate sequencing (highlighted in green). The arrow from ``check condition'' to ``display status'' indicates that display status is next in a sequence, but display status was initially called by ``main'' in a hierarchical call. There is also \errorcode{vague or unclear language}  in the ``word family'' node, since ``determine the pattern were revealed'' could be referring to a variety of functionalities.}
    \label{fig:d38}
\end{figure}

\subsubsection{Illustrative Example 3}\label{sec:results_illustrative_3} In addition to variety in representations, we also found variety in issues that arose. ~\autoref{fig:d38} illustrates a few more: 
\begin{itemize}
    \item  \Node{Node Types} and \Purpose{Purposes}: The nodes in the diagram refer to \nodecode{functions}, each holding a variety of different purposes (labeled in the figure). %The leftmost function is a \purposecode{main function} that is the main program orchestrator and is also used to \purposecode{get player guess}. This node branches out to three nodes. The topmost, ``get guess'', is used to \purposecode{check guess validity}; the one below it, ``word family'', is used to \purposecode{generate word families}, and the bottommost is used to \purposecode{display game status}. To the right of the ``word family'' node is a ``merge patterns'' node, which is used to ``combine the old pattern with the new one to reveal any newly guessed letters'', and above that is a ``check condition'' function. 
    
    \item \Relation{Relation Types}: This diagram uses both relation types. The three relations that branch from the main function, highlighted in red in the diagram, are \relationcode{hierarchical function calls}:
%DZ the caption says that the hierarchical relations are red, not green. Need to fix. %AV Thanks!! Fixed. 
they are all functionality that a main function would use to orchestrate the game. The following relations, highlighted in green in the diagram,
%DZ need to fix relative to caption description %AV Thanks again! Fixed. 
 instead illustrate \relationcode{sequencing} relations. After generating word families, the program updates the current word pattern, checks the win condition, and displays the game status. This clearly lays out a sequence of steps rather than hierarchical calls. % (e.g. ``word family'' does not need to use ``merge patterns'' in order to execute its functionality, but instead is followed by it). 
    
    \item \Error{Issues}: The use of two different relation types is an example of \errorcode{clashing notations}. 
    This mix-and-match is problematic:
    \begin{enumerate}
        \item The use of an arrow is inconsistent, and thus adds ambiguity to the diagram.
        \item Even with a charitable interpretation, it is unclear how a valid program would integrate these two different styles for specifying computation. Attempting to do so can lead to paradoxes; e.g. a back-arrow in a function hierarchy would imply recursion, while in a sequencing diagram, it would imply repetition. 
    \end{enumerate} 
    %Because this sequential portion leads back into the hierarchical portion of the diagram, we now need to concern ourself with sequencing for the hierarchical portion of the diagram even though sequencing was not specified there. 
    %More concretely, the relation between ``check condition'' and ``display status'' tells us that ``check condition'' comes before ``display status'', but in order to incorporate this in the program, we would also need to know where it lies in a sequence with ``get guess'' and ``word family'', which we cannot interpret from the hierarchical calls. 
    %Would the ``merge patterns'' and ``check condition'' functions be called strictly after ``word family''? 

    %Next, we see that (like in the first illustrative example), there is \errorcode{no/implicit game interaction loop}. The diagram implies certain repeated functionality (e.g. using language such as merging an ``old pattern'' with a ``new one'' in the ``merge patterns'' function), but does not explicitly mention looping or repetition.
    
    Finally, we note that there is \errorcode{vague or unclear language} in the ``word family'' node. It is intended to ``get through word list and determine the pattern were revealed. Group the word into families''. In this description, it is not clear what ``determine the pattern were revealed'' is intending to do. Does it determine \textit{if} the pattern is revealed? Or perhaps determine \textit{which} pattern is revealed? It could be repeated functionality with the ``check condition'' node's description of ``check the pattern has no underscore'', but the fact that they are phrased differently (and that this functionality is not required in order to generate word families) makes it unclear if this is what the student intends. 
\end{itemize}

% \begin{figure*}[h]
%     \centering
%     \begin{minipage}[t]{.7\textwidth}
%         \centering
%         \includegraphics[width=0.9\linewidth]{images/D16 Cropped.png}
%         \caption{A diagram with a looser notion of \relationcode{sequencing}. Functions 2 and 3 are used as inputs to Function 4, and which of the two happen first is not critical as long as they both run before Function 4.}
%         \Description{}
%         \label{fig:d16}
%     \end{minipage}
%     \hfill
%     \begin{minipage}[t]{0.25\textwidth}
%         \centering
%         \includegraphics[width=0.7\linewidth]{images/D40 Cropped.png}
%         \caption{A \nodecode{fragment} node.}
%         \Description{}
%         \label{fig:d40}
%     \end{minipage}
% \end{figure*}

%\subsection{Representations} \label{sec:results_representations}

We have now shown most of the representations we found, with full detail included in~\autoref{tab:node-relation-codebook}. Most diagrams were sequencing diagrams (n=32). Several were hierarchical (n=9) and clashing (n=9), and a few were unclear (n=5).  One type of representation that was not shown in our illustrative examples occurred in \textit{fragments}, which included code or pseudocode-like expressions. Students sometimes used fragments to make assignment statements outside of function nodes, with expressions such as ``Guess = Input(``Give a letter'')'', or ``words = [`...'] list of English words'' (quoted from two diagrams not included in the figures shown). Additionally, some sequencing diagrams laid out a well-defined chronological flow (as seen in the second illustrative example in ~\autoref{fig:d51},
%DZ can we put the actual Fig \ref here? %AV added
 with the exception of one relation that created ambiguity). Some diagrams, however, used notions of sequencing in which nodes had multiple incoming or outgoing relations without clear indication of which node should be executed first. Sometimes these relations represented data flow (e.g. two data nodes used as input for a function), and in other cases represented dependency of functionality. Instead of showing a strict ordering of functions, these diagrams showed which functionality or inputs were needed for downstream operations. 

\begin{table}[htbp]
\centering
\small
\renewcommand{\arraystretch}{1.2}
\caption{Node types, purposes, and relations found in student diagrams. Purposes generally arose in functions, with parentheses indicating if they arose in other node types.}
\label{tab:node-relation-codebook}

\arrayrulecolor{lightgray}
\begin{tabular}{
>{\raggedright\bfseries\arraybackslash}p{0.12\linewidth}
>{\itshape\arraybackslash}p{0.24\linewidth}
p{0.55\linewidth}}
\arrayrulecolor{black}
\toprule
\arrayrulecolor{black}
\rowcolor{headergray!80}
\textbf{Dimension} & \textbf{Name} & \textbf{Definition} \\
% \arrayrulecolor{lightgray}
\midrule

% ---------------- Node Type ----------------
\multirow[t]{4}{*}{\Node{Node Type}}
& Function & A function that carries out a purpose. \\
\arrayrulecolor{lightgray}\cline{2-3}\arrayrulecolor{black}
& Fragment & Small portions of functionality that were not enclosed in a function. \\
\arrayrulecolor{lightgray}\cline{2-3}\arrayrulecolor{black}
& Data & Non-function nodes that represent or take in data. \\
\arrayrulecolor{lightgray}\cline{2-3}\arrayrulecolor{black}
%DZ got confused by 'attached other nodes' %AV sorry, added 'to'--is this clear? DZ yes
& Annotation & Auxiliary details or instructions attached to other nodes or relations. \\
\midrule

% ---------------- Node Purpose ----------------
\multirow[t]{20}{*}{\Purpose{Node Purpose}}
& Game loop & Specifies repeating functionality, e.g. using words like ``loop'' or ``until''. \\
\arrayrulecolor{lightgray}\cline{2-3}\arrayrulecolor{black}
& Wrapper function & A function whose only purpose is to call other functions. \\
\arrayrulecolor{lightgray}\cline{2-3}\arrayrulecolor{black}
& Main function & Driver or orchestrator that calls other functions with hierarchical calls. \\
\arrayrulecolor{lightgray}\cline{2-3}\arrayrulecolor{black}
& Initialize number of guesses & Set the number of guesses the player is allowed. \\
\arrayrulecolor{lightgray}\cline{2-3}\arrayrulecolor{black}
& Set word length & Set the length of the word to be guessed. \\
\arrayrulecolor{lightgray}\cline{2-3}\arrayrulecolor{black}
& Initialize list of words & Create the set of all available words for the game. \\
\arrayrulecolor{lightgray}\cline{2-3}\arrayrulecolor{black}
& Filter for words of a given length & Filter words in the word list, e.g. to a certain word length. \\
\arrayrulecolor{lightgray}\cline{2-3}\arrayrulecolor{black}
& Initialize blank word pattern & Create a set of blanks to be filled in by correct guesses. \\
\arrayrulecolor{lightgray}\cline{2-3}\arrayrulecolor{black}
& Generate word families & Generate lists of words grouped by the position of the guessed letter. \\
\arrayrulecolor{lightgray}\cline{2-3}\arrayrulecolor{black}
& Get player guess & Gets the player’s guess as input (function nodes, data nodes). \\
\arrayrulecolor{lightgray}\cline{2-3}\arrayrulecolor{black}
& Check guess validity & Determine if guessed letter is valid, e.g. not repeated, in the alphabet. \\
\arrayrulecolor{lightgray}\cline{2-3}\arrayrulecolor{black}
& Select largest word family & Selects word family with the most words. \\
\arrayrulecolor{lightgray}\cline{2-3}\arrayrulecolor{black}
& Update current word pattern & Update to fill letters in the word pattern with the user guess if needed. \\
\arrayrulecolor{lightgray}\cline{2-3}\arrayrulecolor{black}
& Update guessed letters & After a guess is made, update a list or a total of guessed letters. \\
\arrayrulecolor{lightgray}\cline{2-3}\arrayrulecolor{black}
& Update list of available words & After a guess is made, update the list of available candidate words. \\
\arrayrulecolor{lightgray}\cline{2-3}\arrayrulecolor{black}
& Display game status & Print or display the current game status to the user. \\
\arrayrulecolor{lightgray}\cline{2-3}\arrayrulecolor{black}
& Check win condition & Check if the user has won the game. \\
\arrayrulecolor{lightgray}\cline{2-3}\arrayrulecolor{black}
& Check lose condition & Check if the user has lost the game. \\
\arrayrulecolor{lightgray}\cline{2-3}\arrayrulecolor{black}
& Input data & A node representing input data such as the letter guess, the word list, the number of guesses, or the word length (data nodes). \\
\arrayrulecolor{lightgray}\cline{2-3}\arrayrulecolor{black}
%DZ does 'taking a relation' make sense? %AV it is a sort of figurative use for the word like 'take this road'...I've changed it to "following", do you think that helps? DZ yes
& Condition/control flow & Node specifies conditions or constraints for following a relation to another node (annotation nodes). \\
\arrayrulecolor{lightgray}\cline{2-3}\arrayrulecolor{black}
& Overall game description & Node provides an overall description of the game (annotation nodes). \\
\midrule

% ---------------- Relation Type ----------------
\multirow[t]{2}{*}{\Relation{\textbf{Relation\- Type}}}
& Hierarchical function call & A relation representing a function calling another (helper) function to perform
a subtask. \\
\arrayrulecolor{lightgray}\cline{2-3}\arrayrulecolor{black}
& Sequencing & A relation representing chronological order of execution in which one node is executed before another node. \\
\bottomrule
\end{tabular}
\end{table}

\subsection{Decomposition \& Planning Issues}\label{sec:results_errors}

%nice to start with an easily understandable thing like missing functionality 
We found a variety of issues, with the most common being ~\errorcode{missing functionality}. Almost all diagrams (n=51) were missing required game elements identified by instructors. The number of elements missing in each diagram is detailed in ~\autoref{sec:appendix_elements_and_purposes}. Two 
%DZ I think it's three? 'Looping while the game is not over' was missing in a majority too %AV good catch, thanks! 
elements were missing in a majority of diagrams:  \textit{Updating the pattern so far} (n=39) and \textit{Updating the list of available words} (n=38). Many were also missing the \textit{Looping while the game is not over} element (n=23). This was often missing because no game interaction loop was included (n=19). These 19 also include diagrams that implied repetitive behavior 
%DZ confused here. Are these the remaining 4/23? In Appendix A, there are three needed purposes; game loop, win condition, lose condition... so were these remaining ones missing the win/lose condition or something? %AV yes exactly! The remaining 4 were missing either or both of the win/lose condition--does the sentence I added with the n=4 help? DZ yes. The ones that use 'continuously' without explicitly specifying a loop construct -- are those part of the 19? Or are those ones good/meet the requirements. %AV Those would be part of the 19-- I added "of these", does that make it clearer? 
with language like ``continuously have fewer word families'' without explicitly specifying a loop construct. Others that missed this element did not mention conditions to end the loop (n=4). \autoref{tab:results_error_codebook} summarizes other issues we discovered, which we describe in detail below. Throughout our thematic analysis, we grouped the issues we found into the categories of notations, program quality, encapsulation, clarity, and problem-specific issues. 
\begin{table}[htbp]
\centering
\small
\setlength{\tabcolsep}{4pt}
\renewcommand{\arraystretch}{1.2}
\caption{Issues found in student diagrams.}
\label{tab:results_error_codebook}
%\rowcolors{2}{yellow!20}{white}
\begin{tabular}{
>{\raggedright\bfseries\arraybackslash}p{0.12\linewidth}
>{\raggedright\itshape\arraybackslash}p{0.23\linewidth}
p{0.51\linewidth}
>{\centering\arraybackslash}p{0.06\linewidth}}
\arrayrulecolor{black}
\toprule
\rowcolor{headergray!80}
\textbf{Category} & \textbf{\Error{Issue}} & \textbf{Definition} & \textbf{Count}\\
\midrule

Notations
& Clashing notations
& Relations imply differing notations with incompatible meanings & $9\textcolor{gray}{/55}$\\
\arrayrulecolor{lightgray}\cline{2-4}%\arrayrulecolor{black}
& Non-visual loops in sequencing diagram
& Looping in a sequencing diagram is described in text rather than with visual language & $11\textcolor{gray}{/55}$\\
\arrayrulecolor{lightgray}\cline{2-4}\arrayrulecolor{black}
%& Descriptions without use of nodes
%& Describes several functions' worth of functionality without using visual nodes or relations (N = X) \\
%\arrayrulecolor{lightgray}\cline{2-3}\arrayrulecolor{black}
%\midrule
%\addlinespace[0.35em]
%Order of execution -> notations
& Ambiguous order of execution
& Unclear what order nodes in a sequencing diagram should be visited, e.g. when either of two branching relations could be followed & $ 12\textcolor{gray}{/55}$\\
%& Orphan node - ambiguous
%& Node with no relations; unclear how to reach other nodes (e.g.  ) \\
%\arrayrulecolor{lightgray}\cline{2-3}\arrayrulecolor{black}
%& No/implicit game interaction loop
%& There is no explicit mention of looping or repeating behavior (e.g.  ) \\
\midrule
% \addlinespace[0.35em]
Program Quality
& Hard coded functionality
& Specific data such as constants or letters are embedded (hard coded) in the diagram & $ 15\textcolor{gray}{/55}$\\
\arrayrulecolor{lightgray}\cline{2-4}\arrayrulecolor{black}
& Repeated functionality
& Repeated instructions in separate places that could be achieved by reusing the same functions & $ 14\textcolor{gray}{/55}$\\
\arrayrulecolor{lightgray}\cline{2-4}\arrayrulecolor{black}
& Too many or poorly grouped responsibilities in single function
& A single function has too many purposes and/or purposes that aren't sufficiently related & $ 17\textcolor{gray}{/55}$\\
\arrayrulecolor{lightgray}\cline{2-4}\arrayrulecolor{black}
& Input/output is not a data value
& Input or output of a function is an expression such as a variable assignment or user prompt rather than a data value & $ 11\textcolor{gray}{/55}$\\
\midrule
% \addlinespace[0.35em]
Encapsulation
& Poorly encapsulated loop
& Game functionalities are incorrectly included or excluded in a loop & $ 8\textcolor{gray}{/55}$\\
\arrayrulecolor{lightgray}\cline{2-4}\arrayrulecolor{black}
& Poorly encapsulated function
& Arguments or conditions needed for the function are generated inside the function & $ 10\textcolor{gray}{/55}$\\
\midrule
% \addlinespace[0.35em]
Clarity
& Vague or unclear language
& Underspecified description, or language that is unclear or confusing & $ 13\textcolor{gray}{/55}$\\
\arrayrulecolor{lightgray}\cline{2-4}\arrayrulecolor{black}
& Non-meaningful function name
& Function does not have a meaningful name & $ 17\textcolor{gray}{/55}$\\
\arrayrulecolor{lightgray}\cline{2-4}\arrayrulecolor{black}
& Implementation without clear purpose
& Implementation specifics (e.g. type of loop) without a clear purpose & $ 5\textcolor{gray}{/55}$\\
\midrule
% \addlinespace[0.35em]
Problem-Specific
& Game commits to a single word
& Game commits to a single word rather than updating set of words (misses the `evil trick') & $ 4\textcolor{gray}{/55}$\\
\arrayrulecolor{lightgray}\cline{2-4}\arrayrulecolor{black}
& Selects incorrect word families
& Game selects word families based on an incorrect condition & $ 16\textcolor{gray}{/55}$\\
\arrayrulecolor{lightgray}\cline{2-4}\arrayrulecolor{black}
& Uses a word list instead of\- families
& Directly acts on a word list instead of using word families & $ 4\textcolor{gray}{/55}$\\
\arrayrulecolor{lightgray}\cline{2-4}\arrayrulecolor{black}
& Incorrect order of game operations
& Incorrect order of execution between nodes for correct functionality & $ 3\textcolor{gray}{/55}$\\
\bottomrule
\end{tabular}
\end{table}

% \begin{figure}
%     \centering
%     \includegraphics[width=\linewidth]{images/D38 Annotated.png}
%     \caption{A diagram highlighted with \textit{clashing relation notations} as well as \textit{no/implicit game interaction loop}.}   
%     \label{fig:d38} 
% \end{figure}

\begin{figure}
    \centering
    \includegraphics[width=0.95\linewidth]{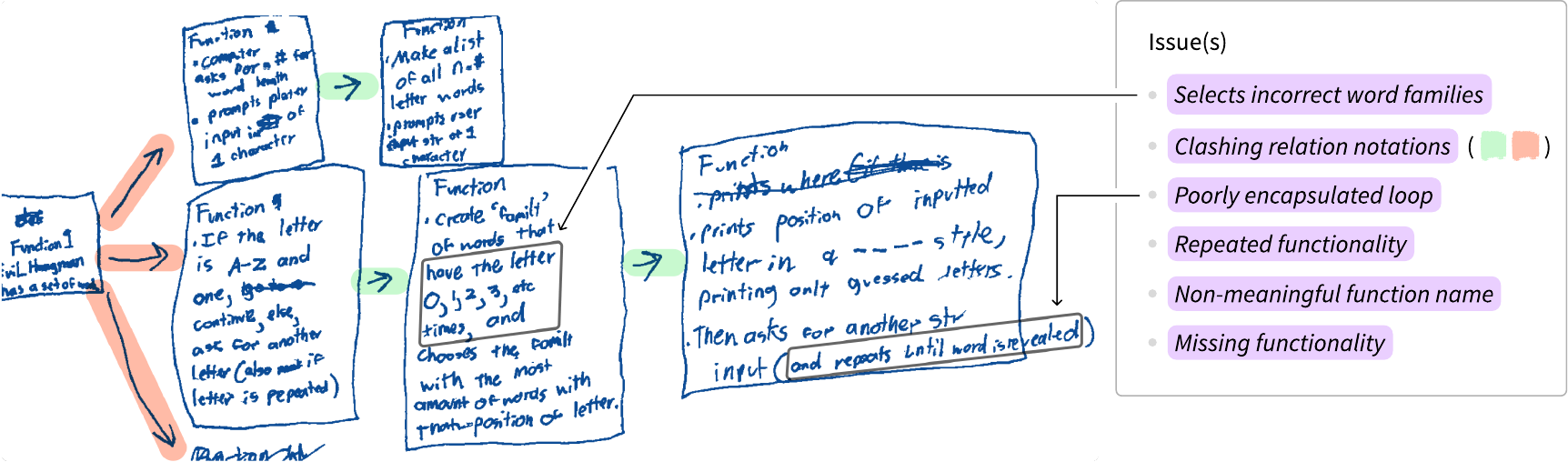}
    %DZ is hierarchical supposed to be red, or green here? %AV hierarchical green, sequencing red. Thanks for checking! DZ but in other figs, hierarchical is red... I thought hierarchical was always red?
    \caption{A diagram with several issues. 1) \errorcode{Selects word families based on incorrect conditions} by filtering for the number of occurrences of the guessed letter. 2) \errorcode{Clashing notations}: the arrows on the left side (red, outgoing from main) indicate a hierarchical function call, while the rest indicate sequencing (green). 3) \errorcode{Poorly encapsulated loop}:  The rightmost node has looping language (``and repeats until word is revealed'')  that presumably only applies to this node, even though purposes in several other nodes should also be repeated. 4) \errorcode{Repeated functionality}: The user is prompted for a guess in four different functions (the two topmost nodes in the diagram and two of the nodes in the middle row). 5) \errorcode{Non-meaningful function name}: Several functions in this diagram are named ``Function'', which is not a descriptive name. 6) \errorcode{Missing functionality}: There are required game elements missing such as updating the current pattern. }
    \label{fig:d36}
\end{figure}

\paragraph{Notations} 
\label{sec:results_clashing}
Issues in notations arose when students used incompatible (clashing) notations or did not meaningfully leverage visual notations. \errorcode{Clashing notations} occurred when one part of the diagram used a sequencing notation and a different part used hierarchical function calls. This mix-and-match adds ambiguity to the diagram: the arrows hold inconsistent meanings and thus make it unclear how the computation should be interpreted or implemented. These can lead to contradictions (e.g. recursion vs. repetition as described in the third illustrative example shown in ~\autoref{fig:d38}). ~\autoref{fig:d36} shows another clash, hierarchical function calls are followed by sequencing. Some diagrams showed a chain of sequencing nodes that ended with a terminal node responsible for looping all functions before it (i.e., acting like a main game loop would in a function call hierarchy), which we also coded as a clashing notation.

%as includes ``Function 1: Evil Hangman'' which makes hierarchical function calls to three helper functions (highlighted in green; bottommost function not included in the figure), and the remaining relations are sequential (highlighted in red). The clashing notations make the overall program execution unclear. 

%\{talk about 3rd example, and then transition from illustrative example to more general}
% This caused issues with interpreting how the functions would be related in a program as described in the third illustrative example in ~\autoref{sec:results_illustrative_examples}. 

%\{move some of the more general text from the 3rd example to here. }
%For example, when does ``Function: Make a list of all $n$ letter words...'' occur with respect to ``Function: create `family' of words that have the letter 0, 1, 2, 3, etc. times...''? Since these occur in different branches of the hierarchical function calls, whose ordering is unspecified, we do not know. 

We also found \errorcode{non-visual loops in sequencing diagrams}, where the game loop is only described in text rather than shown as a relation pointing backwards from a node later in the diagram to an earlier one. % However, some diagrams only described the loop in text. %Any sequencing representation without such a ``backwards'' relation that only described the loop in text would then have this non-visual loops issue. 
For example, one sequencing diagram (not included in the figures) asserted that the game is played ``until the user correctly guesses the solution word'' in a terminal node, without a backwards relation to any previous nodes. 

\begin{figure}
    \centering
    \includegraphics[width=0.95\linewidth]{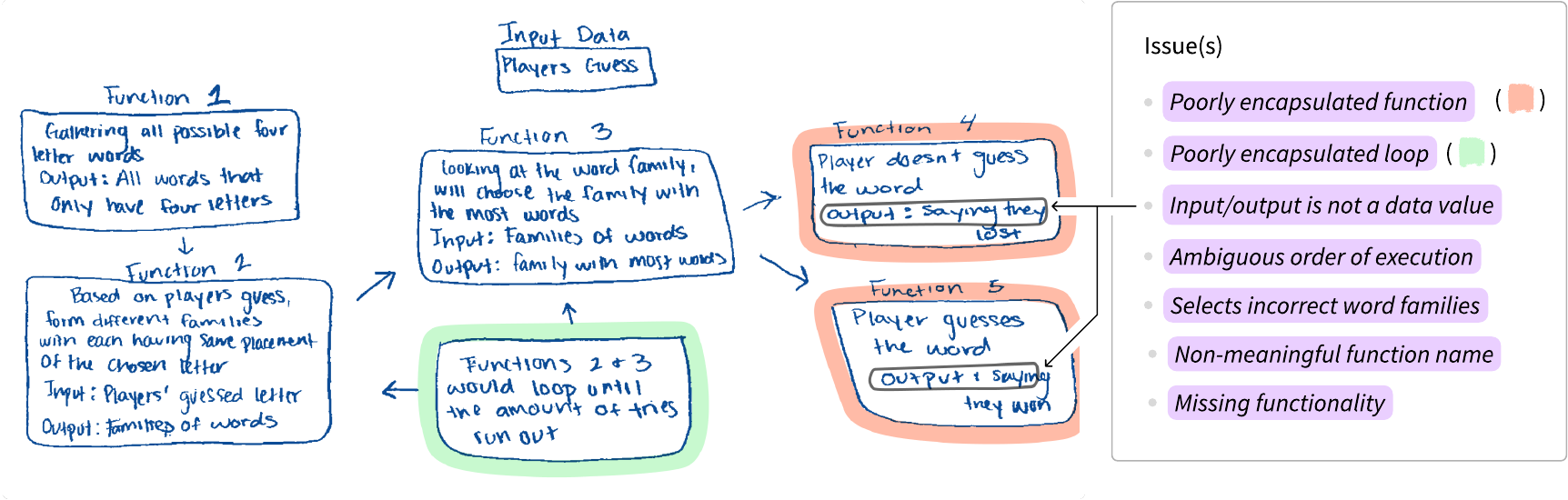}
    %DZ why is 'each having same placement of the chosen letter' wrong? % AV They wrote "Based on players guess, form different families with each having same placement of the chosen letter" it sounds like all the families would have the same placement (whereas what you want is each family to have a different placement of the letter) DZ ahhh, I interpreted it differently, as saying that all words *within* each category have the same letter placement for the guessed letter
    \caption{A diagram with encapsulation and other issues. 1) An annotation node (highlighted in green) states that ``Functions 2 + 3 would loop...'', where Function 2 generates word families and Function 3 selects the largest word family. This is a ~\errorcode{poorly encapsulated loop} because it does not include getting the player's guess, even though it should--the diagram should include the ``Input Data'' node in its loop. 2) There are two \errorcode{poorly encapsulated functions}, Function 4 and Function 5, in which the path to the functions is defined inside of them; the outcome of whether the player guessed the word or not is needed in order to make the branching decision to either Function 4 or Function 5. 3) Functions 4 and 5 have \errorcode{input/output is not a data value}, with ``saying they won'' as an output rather than simply a data value (e.g. a string ``won''). 4) The node at the top with no relations causes \errorcode{ambiguous order of execution}, as it is unclear how to move from this node to the rest of the diagram. 5) It \errorcode{selects incorrect word families} because the word families generated in Function 2 are incorrect in ``each having same placement of the chosen letter''.  6) The functions have \errorcode{non-meaningful function names}, simply labeled with numbers. 7) It has \errorcode{missing functionality}, lacking required game elements including updating the word pattern so far and the list of available words. }
    \label{fig:d25}
\end{figure}

Unclear use of notations can result in an \errorcode{ambiguous order of execution}. For example, some nodes have multiple outgoing branching relations without an indication for when a branch can be taken. Other times, orphan nodes (i.e. nodes with no incoming \& outgoing relations) contributed to the ambiguity (see ~\autoref{fig:d25}).% where it is unclear when each branching relation should be taken.%Another error also reflected a lack of leveraging visual notations: some diagrams had long \errorcode{descriptions without use of nodes}, where several functions' worth of functionality was described in blocks of text rather than being broken into nodes and relations.

% \newpage
% \begin{wrapfigure}[14]{r}{0.4\textwidth}
%   \centering
%     \includegraphics[width=1\linewidth]{images/D32 Cropped.png}
%     \caption{\{Do we want to cut this for space? We already have an ambiguous one in Ex 2}A diagram showing ambiguous order of execution. It is unclear when to move from ``generate\_word\_families()'' to ``update\_hangman()'', and when to move to ``guess\_info()''.}
%     \label{fig:d32_cropped}
% \end{wrapfigure}

%, and ~\autoref{fig:d32_cropped} shows an orphan input data node.

%Alternatively if we want quotes: ``prompts player input...'', ``prompts user input str...'', ``...ask for another letter...'', ``Then asks for another str...''.)
%The text in this diagram specifying repetition (``and repeats until word is revealed'') is presumably referring only to the functionality in this node, even though functionality from several nodes should be repeated

\paragraph{Program quality} There were several issues with  abstraction and reuse. Some diagrams \errorcode{hard coded functionality}, e.g. the number of letters in the word. % (\autoref{fig:d36}).
%DZ is that an issue we should add to that caption? is it annotated in that fig? %AV thanks for catching this--that figure doesn't actually have hard coded functionality, so I took away the figure ref. 
 Other times, diagrams \errorcode{repeated functionality} by failing to abstract and modularize purposes that were executed repeatedly, e.g. multiple functions to get user's guess (\autoref{fig:d36}). Sometimes, there were \errorcode{too many or poorly grouped responsibilities in a single function}, e.g. one diagram (not included in the figures shown) used a single monolithic function with five distinct purposes. Finally, we found diagrams where a function's \errorcode{input/output is not a data value}. These were sometimes egregious, like ``input: loop'' (quoted from a diagram not included in the figures shown), but most described a side-effect instead of a data value, like ``output: saying they won'' (\autoref{fig:d25}). %  An example of non-data outputs is in Functions 4 and 5 of~\autoref{fig:d25}, which have expressions like ``saying they won'' rather than simply a data value such as a string ``won''. This diagram also contains hard coded functionality, as Function 1 selects ``all possible four letter words''. 

\paragraph{Encapsulation} We also find issues with the scope and boundary of a node. A ~\errorcode{poorly encapsulated loop} 
%DZ these are in the reverse order in Table 3. Could we reorder the text here or reorder these two rows in Table 3? %AV agreed, thanks--reordered text to be loop and then function like the table 
occurred when a loop did not have a clear beginning or end, raising questions about which nodes should be repeated (\autoref{fig:d36} and \autoref{fig:d25}). A \errorcode{poorly encapsulated function} occurred when either: (a) the path to the function was defined inside its body (\autoref{fig:d25}), or (b) the function was responsible for producing a value that is passed as an input (e.g. a function responsible for soliciting a guess has the guess as an input parameter). % ~\autoref{fig:d25} shows an example of poorly encapsulated functions as well as a poorly encapsulated loop, and another example of poorly encapsulated looping is in ~\autoref{fig:d36}. 

%  \errorcode{Poorly encapsulated functions} were functions whose inputs (either data needed to execute them, or conditionals needed to determine if they were reached) weren't present until inside the body of the function

%The text in this diagram specifying repetition (``and repeats until word is revealed'') is presumably referring only to the functionality in this node, even though functionality from several nodes should be repeated. }

\paragraph{Clarity} Some diagrams were sparse in meaning by including \errorcode{vague or unclear language}, which made them difficult to parse (\autoref{fig:d38}). Some function nodes had \errorcode{non-meaningful names} (\autoref{fig:d36})
%DZ Fig 8 too %AV I don't think we're aiming to be comprehensive in listing every diagram that contains every error--do you think that's confusing? DZ ahh interesting. I thought in this case it may be odd to not mention it, since the fig does demonstrate this
 which further reduced clarity. We also found \errorcode{implementation without clear purpose}, where diagrams contained blocks of code without supporting commentary on the code's purpose. 

 %AV nothing below here for the Clarity paragraph was included in the initial submission/conditional accept
 %An example of vague or unclear language was described in the third illustrative 
%DZ Fig number
%DZ where did this next line come from? Did I mess this up? I don't think it was there before, but I noticed it in the LaTeX, so I recompiled, and now it's in the pdf too.
%example (~\autoref{fig:d38}). Examples of non-meaningful function names are in many of our figures, such as the several functions named ``Function'' in~\autoref{fig:d36}. %Because this issue was so common (N = X), we did not label it in our figures. 

\paragraph{Problem-specific} Many diagrams contained issues related to the Evil Word Guesser task. %  to solving the problem at hand, often missing the ``evil trick'' or incorrectly narrowing down the word list. 
For example, some students treated the game as the original word-guesser without the ``evil trick,'' i.e., their \errorcode{game commits to a single word} rather than updating a set of words to have as many
%DZ 'many', right? %AV oops, yes! Thanks for the catch! 
 options as possible. Some diagrams \errorcode{select incorrect word families} by, e.g. filtering for words with the fewest instances of a guessed letter rather than word families that leave the most options (\autoref{fig:d36} and \autoref{fig:d25}). Others \errorcode{use a word list instead of families}, precluding correct selection of the smallest \textit{group} of words. Diagrams sometimes had an \errorcode{incorrect order of game operations}, e.g. generating word families before receiving a user guess.% An example of selecting incorrect word families is in ~\autoref{fig:d36}, along with ~\autoref{fig:d25}. 

\section{Discussion}%: Variety and Issues in Representations}
%\todo{Do we want some sort of a brief transition? Take stock of the results?}
We found that some students represent their solutions by communicating ordered sequences of functions, others communicate hierarchies of functions, and some attempt to use these incompatible notations in the same diagram. As we followed each student's solution to interpret their representations of the program at a function level, we found a variety of issues in the ways that they broke down the problem. We now discuss these representational choices and issues, and offer our perspectives about pedagogical implications. 

\subsection{Variety and Issues in Representations}~\label{sec:discussion_variety_and_issues_in_representations}

\paragraph{Students tend towards sequencing} While we instructed primarily using hierarchical representations with some examples of sequencing, we find that the majority of diagrams were sequencing (32 of 55). We also find that students reappropriated hierarchical notations to add sequencing. For example, \autoref{fig:d7} adds sequencing through numbering, such as ``1.'', ``2.'', ``3.'' in the nodes that ``play\_evilhangman()'' points to. Nearly all of the hierarchical diagrams found (8/9) followed reading order; i.e., functions on the left side (or top) of the diagram would be called before those on the right (or bottom). This blend aligns with Sorva's notion of ``higher-level'' notional machines as composites (\autoref{sec:lit_notionalmachines_proceduralOO}). Students use a higher-level notation of hierarchical function calls (an inscription that doesn't explicitly show execution order) but keep track of ordering in ad-hoc ways. This implies they are holding onto two models at once: one based on hierarchical calls and the other based on execution order. While both models have functions as a unit of abstraction, we maintain that sequencing incorporates more low-level detail (specifically execution order) that hierarchical diagrams abstract away. With this framing, hierarchical diagrams represent a higher-level composite notional machine that incorporates both hierarchy and sequencing. Our finding that novices incorporate elements of sequencing also aligns with du Boulay and Lister's work showing that such execution-focused models are commonly used by novice programmers when reasoning about program behavior, particularly in contexts that require explanation, tracing, or planning rather than code production~\cite{duboulay1981blackbox, lister2004multi}. %This (along with Rist's focal expansion model) may provide a fruitful theoretical ground to explore why students tend towards sequencing.

\paragraph{Clashing notional machines} We also observed clashes between hierarchy and sequencing (\autoref{sec:results_clashing}, ~\autoref{fig:d38}, and ~\autoref{fig:d36}). For example, several diagrams included a main function with calls to several sub-functions, which then branched out as sequencing relations (\autoref{fig:d36}). We speculate that student drawings may have begun in the shape of a hierarchical function call --- similar to the diagrams in their textbook (\autoref{fig:textbook_game_decomp}) --- and then diverged to sequencing, closer to a concrete workflow. This shift in abstraction may be explained by Rist's focal expansion model (\autoref{sec:lit_planning_novices}), where novices access familiar schemas in a top-down manner (e.g. the function hierarchy diagram of a game they have seen in their textbook) and resort to a bottom-up process for unfamiliar parts (e.g. specifics of how the user's input is processed).

\paragraph{Challenges with Encapsulation, Abstraction and Reuse} We find that issues traditionally framed as problems in code crop up in diagrams as well. For example, diagrams with poorly encapsulated loops (or missing loops) align with planning errors from past work, such as Spohrer \& Soloway's findings of novices' challenges with identifying boundaries and recognizing how execution of one function affects downstream ones ~\cite{spohrer_soloway_1986novice}. One hypothesis is that these encapsulation issues arise as a result of novices' tendency to focus on local instead of global goals as mentioned in ~\autoref{sec:lit_planning_novices}. If they are focusing solely on a local goal, they may fail to consider how to encapsulate this local goal in such a way that it fits in as a module to enable the global goal of the problem at hand.

Another hypothesis is that novices may not be mentally simulating their solution in depth (or possibly at all), which could lead them not to recognize when there is faulty encapsulation. Without simulating or working through how their solution handles inputs and outputs, students might not recognize ways in which their program has poor encapsulation or program quality practices. For example, let us consider the diagram in ~\autoref{fig:d25} which contains a poorly encapsulated loop (getting the player's guess should be inside the loop, but isn't). Mentally simulating the program represented by this diagram may have led to the realization that taking the player's guess needs to happen in each turn and should therefore be inside the loop. ~\citeauthor{cunningham2017studentsketches}'s work suggests that explicitly tracing through program execution correlates with better performance on programming-related tasks ~\cite{cunningham2017studentsketches}, echoing other work on the affordances of notional machines in helping students understand programs as explained in ~\autoref{sec:lit_notionalmachines}. 

In addition to encapsulation issues, many diagrams also had too many responsibilities assigned to single nodes, hard-coded logic, and limited reuse of functionality. These issues mirror classic violations of modular decomposition principles such as cohesion and information hiding ~\cite{parnas1972criteria, stevens1974structureddesign}. We also find diagrams whose functions had inputs/outputs that were non-data values, like side-effects, which showed that code smells seen in software engineering literature also arise among novices. This suggests that though they are often referred to as ``code quality'' issues, these antipatterns  exist at the planning stage as well. 

\subsection{Implications for Educators and Future Work}
In light of GenAI tools, educators are arguing for the increased importance of teaching decomposition~\cite{porter2024learn, winters2026cs, prather2025beyondthehype}. In this study, we have found that students encounter a number of challenges when performing decomposition, suggesting the need for changes to instruction and avenues for future work.

\paragraph{Representational considerations for educators} The variety and issues in representations we found in student diagrams emphasize the importance of clear instruction in selecting a representation. Educators can consider what attributes they would like their representations to include, and explain the limitations or tradeoffs in their chosen representation. We instructed primarily using hierarchical representations and with some examples of sequencing, and many students chose the latter (n=32) or incorporated elements of both  (n=9). This highlights that different representations afford different kinds of reasoning while obscuring others. Students repeatedly attempted to encode order of execution, suggesting a tendency towards procedural reasoning of program behavior. For the representations instructors choose, examples and explanations of features of a notation could help students more effectively use them. A relevant instructional method is use of multiple representations to show different approaches to the same problem~\cite{ainsworth1999functions, ainsworth2006deft}. For example, an instructor could show one hierarchical solution and one sequencing solution to a task. These comparisons can help students understand how to use a representation and the kinds of affordances the representation provides. Instructors may also consider ways to mitigate confusion when the same symbol has different meanings (e.g. an arrow can represent either sequencing or a function call). They may use color or add annotations for the different types of nodes or relations to clarify the role of each element \cite{zhang2026notationsevolvehistoricalanalysis}. % Making these potential confusions explicit could alert students to possible issues that may arise as they decompose problems. 
%AV added back in after conditional accept
\paragraph{\Yellow{Recommendation}}Educators should consider explicitly teaching the affordances and limitations of their chosen notation to help students maintain a consistent mental model when reasoning about sequencing, data flow, and function call structures. 
%DZ cut -- trying to make these one-sentence takeaways. For me, the key here is that each decomposition notation has tradeoffs. 
%Showing contrasting examples of the same solution in different representations can help make these differences clear.  For similar and easily conflatable components, devise notations to create clear differences (e.g. different colors for data vs. function nodes or hierarchical vs. sequencing arrows). 

%LP - this next paragraph has two key ideas and I fear the latter idea is being lost by being too late in the paragraph.  If you split this into two paragraphs starting at "A challenge in teaching decomposition", it allows for this second point to be a core, concrete recommendation.  It'd  let the title of the section for this paragraph be meaningful and would allow for expansion on this point.  A title for this new section could be "Diagrams facilitate easier feedback on decomposition". DZ: we split this out.

%LP - the first sentence in the next paragraph is a bit wild.  What instructor doesn't teach inputs, outputs, and purpose of functions when first introducing them?   That doesn't seem like a unique contribution of this work. DZ: rephrased this around specification and function design
\paragraph{Structured decomposition instruction} Educators can also consider decomposition and planning-focused methods of instruction and assessment. We encourage educators to use a structured function design process, where students describe example cases, parameter and return types, and function descriptions, all \textit{before} moving toward implementation.
This may help with some of the issues we found in student diagrams around encapsulation and inputs and outputs that are not data values. One example of instructional material that emphasizes modularity is ``How to Design Programs'' ~\cite{felleisen2018design}, which encourages students to identify a function's ``contract and purpose'' (inputs, outputs, and purpose) before implementation. 
\paragraph{\Yellow{Recommendation}} Educators should consider introducing a structured function design approach that delineates specification from implementation.

\paragraph{Diagrams facilitate rapid assessment and feedback on decomposition} 
In the past, CS1 instructors have generally required students to submit fully working software for their assignments. For decomposition to be assessed at all, it would therefore need to be reverse-engineered from notions of quality evident in the code. We encourage educators to assess decomposition as its own skill, distinct from (possible) future implementation. In particular, diagrams may provide a lightweight approach for assessing decomposition many times throughout the course, disconnected from the limited number of ``full'' programming projects that can be assigned.

%LP - the recommendation paragraphs are redundant with the text above.  We could either remove the redundancy or remove the "Recommendation" sections. DZ: keeping them but tightened them up

\paragraph{\Yellow{Recommendation}} Educators may consider assessing decomposition as its own bona fide skill, distinct from implementation projects.

\paragraph{Future work in higher-level plan tracing practices among novices} Educators and researchers interested in decomposition may consider studies where students simulate the plans they create by tracing through their solutions. A key affordance of notional machines in pedagogical contexts is simulation~\cite{sorva2013, duboulay1981blackbox,sajaniemi2008procedures}. Tracing in introductory computing has been studied extensively at the level of small portions of code, such as in work by ~\citeauthor{lister2004multi} and ~\citeauthor{cunningham2017studentsketches}~\cite{lister2004multi, cunningham2017studentsketches}. Decomposition diagrams offer a way for students to trace through program plans at a high level, following relationships between functions rather than individual lines of code. Future work could examine whether students who simulate the program described by their diagram can catch some of the issues that we found. For example, after drawing a diagram with an ambiguous order of execution, perhaps students recognize this ambiguity as they try to decide which of two branching relations to follow. Similarly, simulation may help students recognize encapsulation issues as discussed in ~\autoref{sec:discussion_variety_and_issues_in_representations}. Studies could investigate novices' simulations of these higher-level representations of programming plans (which we might call \textit{plan tracing}). Researchers conducting these studies may consider observing the participants' process in addition to artifacts like our students' diagrams. These studies could offer insights into how students simulate solutions, and the order in which they approach parts of a problem. 
\paragraph{\Yellow{Recommendation}} Future studies may explore how prompting students to simulate their solutions affects the quality of their decomposition. We may consider exploring \textit{plan tracing} as an emerging computing competency and skill.

%Overall, give ~1 sentence examples for what you could do
%- Showing multiple approaches and pros/cons --"Choose a representation that emphasizes desired attributes, and explain how to handle limitations of the chosen representation." Trim/combine this subsection to be 1 paragraph. include the part about showcasing tradeoffs. AV: maybe also mention simulation here. 
% - A path might be to teach decomp earlier -- "Module-first Decomposition Instruction" (cut starting at one example or after 1 sentence on HtDP as an example)

% 5.3 Future work %% 1 paragraph
% - possible next steps are:
%   -- is think-alouds
%   -- connecting diagrams to composition
%   -- having students trace through their diagrams to find disconnects
%   -- seeing how advanced students/professionals approach the diagrams

\section{Conclusion}

We characterized representations and issues in CS1 students' problem decompositions when given a design task and a blank page. Our results show that students used varied representations and reappropriations of hierarchical and sequencing notations. We also find issues involving notation, execution order, abstraction and reuse, encapsulation, clarity, and problem-specific misunderstandings. Our work builds on prior planning and decomposition literature, contributing knowledge of how novices decompose complex problems in unconstrained environments. With more educators now advocating for teaching decomposition at the introductory level, our work provides timely insights into common pitfalls in diagramming and program design. We hope this motivates future research on supporting students' learning of core decomposition skills, by both helping students better represent their decomposition (e.g. using more meaningful notations), and by helping instructors explicitly teach decomposition and diagramming (e.g. by justifying their chosen decomposition notations to students).

\begin{acks}
We thank Alex Chao, Bill Griswold, and our anonymous reviewers, for their helpful feedback. 

This work is supported by the National Science Foundation under Grant No. 2417374, the Google Award for Inclusion Research, the Microsoft AI Economy Institute, and the GenAI in CS Education Consortium. Any opinions, findings, and conclusions or recommendations expressed in this material are those of the author(s) and do not necessarily reflect the views of other entities.
\end{acks}

\appendix 

\section{Appendix: Required Game Elements and Purposes}\label{sec:appendix_elements_and_purposes}

Our analysis of the purposes of each node in the diagrams yielded a variety of codes (enumerated in ~\autoref{tab:node-relation-codebook}). Purposes that correspond with the required game elements are in the middle
%DZ middle now, right?
 column of \autoref{tab:game_elements_purposes_codes} (parentheses indicate subcodes). 

\begin{table}[h]
\centering
\caption{Required game elements from instructors, purposes of nodes identified through qualitative coding, and number of diagrams \textbf{missing} each element.}
\label{tab:game_elements_purposes_codes}
\small
\renewcommand{\arraystretch}{1.2}
\rowcolors{2}{gray!6}{white}
\begin{tabularx}{\columnwidth}{p{0.35\columnwidth}p{0.5\columnwidth}p{0.08\columnwidth}}
\toprule
\textbf{Required Game Elements From Instructors} &
\textbf{Corresponding Purposes} & 
\textbf{Count Missing} \\ 
\midrule
Initializing list of words of the chosen length
& Filter for words of a given length; Set word length; Initialize list of words (words of a given length); Input data (word list) 
& 11\textcolor{gray}{/55} \\
%\midrule
Getting next letter guess from user
& Get player guess; Input data (letter guess) \
& 7\textcolor{gray}{/55} \\
%\midrule
Creating word families from the list of available words
& Generate word families 
& 14\textcolor{gray}{/55} \\
%\midrule
Selecting the largest word family
& Select largest word family 
& 19\textcolor{gray}{/55} \\
%\midrule
Updating the pattern so far
& Update current word pattern 
& 39\textcolor{gray}{/55}\\
%\midrule
Updating the list of available words
& Update list of available words 
& 38\textcolor{gray}{/55} \\
%\midrule
Looping while the game is not over
& Game loop; Check win condition; Check lose condition 
& 23\textcolor{gray}{/55}\\
\bottomrule
\end{tabularx}
\end{table}

Our bottom-up coding process often yielded codes that did not have a one-to-one mapping with the required game elements. For example, ``Looping while the game is not over'' did not emerge from our coding, although the three purposes listed in the middle column did; to meet this requirement, a student's solution would have needed to support these three purposes across one or more nodes.
%DZ please check wording above
In other cases, such as ``Initializing list of words of the chosen length'', not all of the codes shown are needed; a student may, for example, have a diagram that separately takes in ``Input data (word list)'', then does ``Set word length'', and then does ``Filter for words of a given length'' on the word list. A different student may have simply had ``Initialize list of words (words of a given length)'' as a single step. Both of these approaches would have satisfied this requirement. 

%DZ OK to delete this for space? Repetitive with above + beginning of Sec 4.2.
%For ``Looping while the game is not over'', this element was considered missing if a diagram was missing any of the three corresponding purposes (that is, in order for this element to be considered present, the diagram would need to have all three of the corresponding purposes). Of the 23 diagrams missing this element, 19 did not explicitly mention or show a game loop (no `Game loop' purpose was coded). 

\bibliographystyle{ACM-Reference-Format}
\bibliography{refs}

\end{document}